\documentclass{article}

\usepackage{algorithm}  
\usepackage{algpseudocode}
\usepackage{amsmath}
\usepackage{amssymb}
\usepackage{authblk}
\usepackage{bbold}
\usepackage{caption}
\usepackage{doi}
\usepackage[left=35mm, right=35mm, top=32mm, bottom=32mm]{geometry}
\usepackage{graphicx} 
\usepackage{hyperref}
\usepackage[natbib=true, style=authoryear, giveninits=true, terseinits=true, maxcitenames=2, maxbibnames=10, uniquename=mininit, uniquelist=false, dashed=false]{biblatex}
\usepackage{titlesec}
\usepackage{xfrac}

\hypersetup{
    colorlinks=True, 
    allcolors=blue
}

\bibliography{references}

\DeclareNameAlias{sortname}{family-given}
\DefineBibliographyStrings{english}{%
    andothers={\textit{et al}\adddot},
    backrefpage={page},
    backrefpages={pages},
    edition={Edition},
    mathesis={Master's thesis},
    references={\normalsize\MakeUppercase{References}},
}

\DeclareFieldFormat{pages}{#1}
\DeclareFieldFormat[article, inbook, inproceedings, thesis]{title}{#1} 
\DeclareFieldFormat[article]{volume}{\mkbibbold{#1}}
\AtEveryBibitem{\clearfield{number}}
\renewbibmacro{in:}{}

\AtBeginBibliography{%
}

\newcommand{\norm}[2][]{\Vert #2\Vert_{#1}}
\newcommand{\bs}[1]{\ensuremath{\boldsymbol{#1}}}

\newcommand{\data}{y}
\newcommand{\bdata}{\bs{\data}}
\newcommand{\bdataO}{\bdata^{\mathrm{obs}}}
\newcommand{\datadim}{d}
\newcommand{\error}{e}
\newcommand{\berror}{\bs{\error}}
\newcommand{\pred}{p}
\newcommand{\bpred}{\bs{\pred}}
\newcommand{\preddim}{m}
\newcommand{\param}{k}
\newcommand{\bparam}{\bs{\param}}
\newcommand{\paramdim}{n}

\newcommand{\dd}{\mathrm{d}}

\newcommand{\fwd}{f}
\newcommand{\bfwd}{\bs{\fwd}}
\newcommand{\Fwd}{\bs{F}}
\newcommand{\predmodel}{q}
\newcommand{\bpredmodel}{\bs{\predmodel}}
\newcommand{\Predmodel}{\bs{Q}}

\newcommand{\numsamples}{J}

\newcommand{\lengthscale}{\ell}

\newcommand{\cov}{\bs{\Gamma}}
\newcommand{\chol}{L}
\newcommand{\bchol}{\bs{\chol}}

\newcommand{\x}{\bs{x}}

\newcommand{\CompNum}{\gamma}
\newcommand{\Perm}{\kappa}
\newcommand{\Pressure}{P}

\title{\itshape\Large Data Space Inversion for Efficient Predictions and Uncertainty Quantification for Geothermal Models}

\author[1]{Alex~de~Beer\thanks{Corresponding author. Email: \href{mailto:adeb0907@uni.sydney.edu.au}{adeb0907@uni.sydney.edu.au}.}}
\author[2]{Andrew~Power}
\author[2]{Daniel~Wong}
\author[2]{Ken~Dekkers}
\author[2]{Michael~Gravatt}
\author[3]{Elvar~K.~Bjarkason}
\author[2]{John~P.~O'Sullivan}
\author[2]{Michael~J.~O'Sullivan}
\author[2]{Oliver~J.~Maclaren}
\author[2]{Ruanui~Nicholson}

\affil[1]{School~of~Mathematics~and~Statistics,~University~of~Sydney,~New~South~Wales~\emph{2006},~Australia}
\affil[2]{Department~of~Engineering~Science~and~Biomedical~Engineering,~University~of~Auckland, Auckland~\emph{1010},~New~Zealand}
\affil[3]{Graduate~School~of~International~Resource~Sciences, Akita~University,~Akita~\emph{010-8502},~Japan}

\titleformat{\section}{\normalsize\bfseries\uppercase}{\thesection}{1em}{}
\titleformat{\subsection}{\normalsize\bfseries}{\thesubsection}{1em}{}
\titleformat{\subsubsection}{\normalsize\itshape}{\thesubsubsection}{1em}{}

\newenvironment{abs}{
    \small
    \begin{center}
        {\bfseries\MakeUppercase{\abstractname}}
    \end{center}
    \quotation
}
{\endquotation}

\date{}

\begin{document}

\maketitle

\begin{abs}
    The ability to make accurate predictions with quantified uncertainty provides a crucial foundation for the successful management of a geothermal reservoir. Conventional approaches for making predictions using geothermal reservoir models involve estimating unknown model parameters using field data, then propagating the uncertainty in these estimates through to the predictive quantities of interest. However, the unknown parameters are not always of direct interest; instead, the predictions are of primary importance. Data space inversion (DSI) is an alternative methodology that allows for the efficient estimation of predictive quantities of interest, with quantified uncertainty, that avoids the need to estimate model parameters entirely. In this paper, we illustrate the applicability of DSI to geothermal reservoir modelling. We first review the processes of model calibration, prediction and uncertainty quantification from a Bayesian perspective, and introduce data space inversion as a simple, efficient technique for approximating the posterior predictive distribution. We then introduce a modification of the typical DSI algorithm that allows us to sample directly and efficiently from the DSI approximation to the posterior predictive distribution, and apply the algorithm to two model problems in geothermal reservoir modelling. We evaluate the accuracy and efficiency of our DSI algorithm relative to other common methods for uncertainty quantification and study how the number of reservoir model simulations affects the resulting approximation to the posterior predictive distribution. Our results demonstrate that data space inversion is a robust and efficient technique for making predictions with quantified uncertainty using geothermal reservoir models, providing a useful alternative to more conventional approaches.
\end{abs}

\section{Introduction}

Computational models are widely used in geothermal reservoir engineering to facilitate effective decision making \citep{osullivan2016reservoir}. One of the key features of these models is the ability to make predictions with quantified uncertainty. Computing accurate predictions and uncertainty estimates generally requires calibration of the model; that is, the estimation of model parameters, such as the subsurface permeability structure and the strength and location of the deep mass upflows at the base of the system, using observations such as downhole temperature and pressure measurements. In many situations, the parameters themselves are not of direct interest; instead, the predictions are of primary importance. However, the calibration process is typically the most computationally demanding step in the process of making predictions. 

Here we discuss the application of the data space inversion (DSI) methodology \citep{sun2017new, sun2017production} for making predictions using geothermal reservoir models, as well as providing associated estimates of uncertainty. The DSI framework provides several computational advantages, the most significant being the ability to effectively circumvent the model calibration process; instead, it simply estimates the values of predictive quantities of interest conditioned on measured data. This idea is sometimes referred to as {\em direct forecasting}. Furthermore, the DSI approach does not require access to model derivatives (in fact, the method can be applied to non-differentiable models) while much of the required computation can be carried out in parallel. The DSI approach has been used successfully in a variety of applications, including subsurface hydrology \citep{delottier2023data}, petroleum engineering \citep{jiang2020data, jiang2021treatment, lima2020data, liu2021improved}, and carbon storage \citep{sun2019data, jiang2024history}. The application of the framework to geothermal reservoir modelling, however, is largely unexplored and potentially more challenging as the governing equations are generally highly nonlinear in the geothermal context, with simulation non-convergence being a common issue \citep[see, e.g.,][]{croucher2020waiwera, o2013improved}.

We note that the idea of direct forecasting is not exclusive to DSI; in particular, the Bayesian evidential learning (BEL) framework \citep{scheidt2018quantifying} also involves direct forecasting. Like DSI, the BEL framework has been applied in a variety of subsurface modelling applications \citep{hermans2018uncertainty, michel20201d, pradhan2020seismic}; most notably, \citet{athens2019monte} demonstrate the application of BEL to predict the temperature in a geothermal target area of a synthetic model based on Dixie Valley, Nevada. We note, however, that only single-phase, natural state simulations are considered in this study; by contrast, we apply the DSI framework to a two-phase problem, and consider both natural state and production history simulations.

In this paper, we build on our previous work \citep{power2022data} to illustrate the applicability of the DSI methodology to geothermal reservoir modelling. The present work makes a number of new contributions to the DSI literature. First, we introduce a modification of the typical DSI algorithm that allows us to sample from the resulting approximation to the posterior predictive distribution directly and efficiently, avoiding the need to characterise this distribution using techniques such as randomised maximum likelihood \citep{sun2017new, sun2017production} or ensemble methods \citep{lima2020data}, which add an additional degree of approximation to the procedure. Second, we illustrate the applicability of the resulting DSI algorithm to geothermal reservoir modelling through two synthetic model problems (outlined in Section \ref{sec:examples}); one based on a simplified two-dimensional reservoir and one based on a large-scale, three-dimensional reservoir. Through these model problems, we provide a numerical comparison between DSI and other methods for uncertainty quantification in subsurface modelling, investigate how the number of reservoir model simulations affects the resulting approximation to the posterior predictive distribution, and illustrate how one can evaluate the accuracy of the approximate mapping constructed during the DSI algorithm.

\section{Methodology}

In this section, we first introduce some key notation used throughout the remainder of the paper. We then briefly recall the key concepts and steps involved in a typical (Bayesian) statistical approach to geothermal model calibration and prediction, and the associated uncertainty quantification. Finally, we introduce data space inversion as a technique that allows us to circumvent some of the difficulties associated with classical approaches to model calibration and prediction in the Bayesian setting.

\subsection{Notation}

Throughout the paper, we use bold lowercase letters to denote vectors and bold uppercase letters to denote matrices. We use $\bs{I}_{n}$ to denote the identity matrix of dimension $n$, and $\bs{0}_{n}$ to denote the zero vector of dimension $n$. We use the notation $\{x^{(i)}\}_{i=1}^{n}$ as a shorthand for the set $\{x^{(1)}, x^{(2)}, \dots, x^{(n)}\}$. For a symmetric positive definite matrix $\bs{G}\in\mathbb{R}^{n\times n}$ and vector $\bs{v}\in\mathbb{R}^n$, we let $\norm[\bs{G}]{\bs{v}} := \sqrt{\bs{v}^{\top}\bs{G}\bs{v}}$ denote the Euclidean norm weighted by $\bs{G}$. Finally, we use the notation $\bs{u} \sim \mathcal{N}(\bs{u}_{0}, \bs{\cov})$ to indicate that a random variable, $\bs{u} \in \mathbb{R}^{n}$, has a Gaussian distribution with mean $\bs{u}_{0} \in \mathbb{R}^{n}$ and covariance matrix $\bs{\cov}\in\mathbb{R}^{n\times n}$.

\subsection{Bayesian Model Calibration and Prediction}

Uncertainty quantification (UQ) for geothermal reservoir models typically uses the Bayesian framework~\citep{kaipio2006statistical, tarantola2005inverse}, which naturally allows for incorporation and quantification of various sources and types of uncertainty. The standard procedure is as follows:
\begin{enumerate}
    \item Calibrate (i.e., estimate) the model parameters using data. 
    \item Approximately (i.e., linearly) quantify the posterior uncertainty in the parameters. 
    \item Propagate the posterior parameter uncertainty to the predictive quantities of interest. 
\end{enumerate}
We now describe each of these steps in greater detail.

\subsubsection{Calibration} \label{sec:calibration}

Calibration of geothermal reservoir models is typically based on the assumption that the parameters and data are linked by a setup of the form
\begin{align}
    \bdata = \bfwd(\bparam) + \berror,
\end{align}
where $\bdata\in\mathbb{R}^\datadim$ denotes the data, $\bparam\in\mathbb{R}^\paramdim$ denotes the unknown parameters, $\bfwd:\mathbb{R}^\paramdim\rightarrow\mathbb{R}^\datadim$ represents the forward model (or \emph{parameter to observable} mapping), and $\berror\in\mathbb{R}^\datadim$ denotes measurement error (as well as possible model error). Generally, application of the forward model $\bfwd(\cdot)$ involves first solving for the the dynamics of the system using a reservoir simulator (for further details, see Section \ref{sec:equations}), then applying an observation operator which extracts the simulation output at the times and locations corresponding to the available data.

In a Bayesian setting, the model calibration problem is recast as a problem of statistical inference, where the goal is to estimate the (parameter) posterior distribution, $\pi(\bparam\vert \bdataO)$; that is, the conditional distribution of the parameters given a particular realisation of the data, $\bdataO$. The posterior distribution can be expressed using Bayes' theorem, as
\begin{align}
    \pi(\bparam\vert\bdataO) = \frac{\pi(\bdataO\vert\bparam)\pi(\bparam)}{\pi(\bdataO)}\propto\pi(\bdataO\vert\bparam)\pi(\bparam). \label{eq:bayes}
\end{align}
In Equation \eqref{eq:bayes}, $\pi(\bparam)$ denotes the \emph{prior distribution}, which describes our beliefs about the parameters before (i.e., prior to) considering the data, while the \emph{likelihood}, $\pi(\bdataO|\bparam)$, encodes the likelihood of observing the data, $\bdataO$, under a given set of parameters. The marginal probability $\pi(\bdataO)$ acts as a normalising constant, and is unimportant in most cases.

Assuming the errors, $\berror$, are independent of the parameters, $\bparam$, it is well known that the likelihood inherits the distribution of the errors; that is, $\pi(\bdataO\vert\bparam)=\pi_e (\bdataO-\bfwd(\bparam))$ \citep[see, e.g.,][]{calvetti2007introduction, kaipio2006statistical}. We can therefore rewrite Equation \eqref{eq:bayes} as
\begin{align}
    \pi(\bparam\vert \bdataO)\propto\pi(\bdataO\vert\bparam)\pi(\bparam)=\pi_e (\bdataO-\bfwd(\bparam))\pi(\bparam).
\end{align}
Assuming a-priori, as is standard, that the parameters and errors are normally distributed---that is, $\bparam \sim \mathcal{N}(\bparam_0, \cov_{\param})$ and $\berror\sim\mathcal{N}(\bs{0}_{\datadim},\cov_{\error})$---the posterior can be expressed as~\citep{calvetti2007introduction, kaipio2006statistical}
\begin{align}
    \pi(\bparam\vert \bdataO)\propto\exp\left\{-\frac{1}{2}\left(\norm[\cov_{\error}^{-1}]{\bdataO-\bfwd(\bparam)}^2+\norm[\cov_{\param}^{-1}]{\bparam-\bparam_0}^2\right)\right\}.
\end{align}
Computing a full characterisation of the posterior generally requires the use of sampling-based approaches such as Markov chain Monte Carlo (MCMC) \citep{cui2011bayesian, cui2019posteriori, maclaren2020incorporating, scott2022bayesian}. However, for most geothermal models these methods are computationally infeasible. As a computationally cheaper alternative, it is common to compute the {\em maximum a posteriori} (MAP) estimate, $\bparam_{\mathrm{MAP}}$; that is, the point in parameter space which maximises the posterior density, defined as
\begin{equation}
    \bparam_{\mathrm{MAP}} := \arg \min_{\bparam\in\mathbb{R}^{\paramdim}} \left\{ \frac{1}{2}\norm[\cov_{\error}^{-1}]{\bdataO-\bfwd(\bparam)}^2+\frac{1}{2}\norm[\cov_{\param}^{-1}]{\bparam-\bparam_{0}}^2 \right\}. \label{eq:map_opt}
\end{equation}
The definition of the MAP estimate reveals a link between the Bayesian approach to model calibration and classical, optimisation-based approaches; computing the MAP estimate is equivalent to solving a regularised least-squares problem, where the form of the regularisation term follows from the specification of the prior. 

\subsubsection{Approximate Parameter Uncertainty Quantification} \label{sec:local_lin}

After computing the MAP estimate, it is common to quantify the posterior uncertainty approximately, using a local Gaussian approximation~\citep[see, e.g.,][]{omagbon2021case, zhang2014inference}; that is, $\pi(\bparam\vert\bdataO)\approx\mathcal{N}(\bparam_{\rm MAP},\cov_{\mathrm{post}})$. A common approximation to the posterior covariance matrix is given by
\begin{align}
    \cov_{\mathrm{post}} = (\Fwd^\top\cov_{\error}^{-1}\Fwd+\cov_{\param}^{-1})^{-1}, \label{eq:local_lin}
\end{align}
where $\Fwd$ denotes the sensitivity (Jacobian) matrix of the model with respect to the parameters (i.e., $\Fwd_{ij}=\sfrac{\partial \fwd_i}{\partial \param_{j}}$, for $i = 1, 2, \dots, \datadim$ and $j = 1, 2, \dots, \paramdim$), evaluated at the MAP estimate. A sample, $\bparam_{i}$, from the Gaussian approximation to the posterior, $\mathcal{N}(\bparam_{\rm MAP},\cov_{\rm post})$ can be generated as 
\begin{align}\label{eq: samps}
    \bparam_{i} = \bparam_{\rm MAP} + \bchol_{\rm post}\bs{\eta}_{i},
\end{align}
where $\bchol_{\rm post}\bchol_{\rm post}^\top=\cov_{\rm post}$, and $\bs{\eta}_{i} \sim \mathcal{N}(\bs{0}_{\paramdim},\bs{I}_{\paramdim})$ is a sample from the $\paramdim$-dimensional standard Gaussian distribution. Because this approximation uses a linearisation of the forward model, it is often referred to as \emph{linearisation about the MAP estimate} (LMAP).

We note that there exist several additional classes of methods that have been used to approximate the parameter posterior and posterior predictive distributions in geothermal settings. One such method is randomised maximum likelihood \citep[RML;][]{kitanidis1995quasi, oliver1996conditioning}, in which a set of stochastic optimisation problems are solved to obtain samples distributed in regions of high posterior density \citep[geothermal applications include][]{bjarkason2020uncertainty, tian2024advanced, tureyen2014assessment, zhang2014inference}. These optimisation problems take the same form as Equation \eqref{eq:map_opt}, but the data is perturbed using a sample from the distribution of the error and the prior mean is replaced by a sample from the prior. While RML generally produces a more accurate approximation to the posterior than linearisation about the MAP estimate, the computational expense is amplified given that each sample from the (approximate) posterior incurs a similar computational cost to computing the MAP estimate. An alternative class of methods are ensemble methods \citep{chen2013levenberg, emerick2013ensemble, iglesias2013ensemble}, which are based on sampling methods such as RML, but employ an ensemble (sample-based) approximation of any required derivative or covariance information \citep[geothermal applications include][]{bekesi2020updated, bjarkason2020uncertainty}. These are simpler to implement and can be less computationally intensive than methods such as RML. However, the calibration process is still iterative and potentially requires many forward simulations.

\subsubsection{Approximate Predictive Uncertainty Quantification} \label{sec:pred_uq_approx}

Carrying out uncertainty quantification for the predictive quantities of interest, denoted here by $\bpred\in\mathbb{R}^{\preddim}$, relies on having a \emph{predictive model}, $\bpredmodel : \mathbb{R}^{\paramdim}\rightarrow\mathbb{R}^{\preddim}$, relating the parameters to the predictions; that is,
\begin{align}
    \bpred = \bpredmodel(\bparam).
\end{align}
In various settings, the forward model $\bfwd(\cdot)$ and predictive model $\bpredmodel(\cdot)$ may be represented by the same model but evaluated at different locations in space and/or time. In any case, to propagate the uncertainty from the parameters through to the predictions, typically one of two methods is applied \citep{omagbon2021case}:
\begin{enumerate}
    \item Samples from the (approximate) parameter posterior are generated (see Equation~\eqref{eq: samps}) and run through $\bpredmodel$ to give prediction samples; that is, $\bpredmodel(\bparam_{i})$, where $\bparam_{i}\sim\mathcal{N}(\bparam_{\rm MAP}, \cov_{\rm post})$.
    \item A Gaussian approximation of the posterior predictive distribution is made; that is, $\bpred\sim\mathcal{N}(\bpred_{\rm MAP},\cov_{\rm pred})$, with 
    \begin{align}
        \bparam_{\rm MAP} = \bpredmodel(\bparam_{\rm MAP}),\quad 
        \cov_{\rm pred} = \Predmodel\cov_{\rm post}\Predmodel^\top,
    \end{align}
    where $\Predmodel$ denotes the sensitivity (Jacobian) matrix of the predictive model with respect to parameters (i.e., $\Predmodel_{ij} = \sfrac{\partial \predmodel_{i}}{\partial \param_{j}}$) for $i = 1, 2, \dots, \preddim$ and $j = 1, 2, \dots, \paramdim$, evaluated at the MAP estimate.
\end{enumerate}
The latter approach is often termed {\em linear(-ised) uncertainty propagation}, and may not be suitable in all situations; see \citet{omagbon2021case} for a comparison and discussion of these two approaches within a geothermal context. We emphasise that both of the aforementioned approaches only approximate the posterior predictive distribution (with the exception of the case where both the forward and predictive models are linear and the prior and error distributions are Gaussian).

\subsubsection{Accurate Predictive Uncertainty Quantification} \label{sec:pred_uq}

Although infeasible in most geothermal settings, it is theoretically possible to accurately characterise the posterior predictive distribution. Specifically, this involves first generating samples from the true posterior using (for example) MCMC, then running these samples through the predictive model. In this work, we use this approach to provide a benchmark for the simplified two-dimensional reservoir (see Section~\ref{sec:Darcy}).

\subsection{Data Space Inversion} \label{sec:dsi}

The main computational bottleneck associated with the standard procedure for (approximate) uncertainty quantification is computation of the MAP estimate, $\bparam_{\rm MAP}$. Due to the scale and computational complexity of a typical geothermal model, solving the optimisation problem \eqref{eq:map_opt} can require a significant amount of time, even when efficient methods such as adjoint-based approaches are used \citep{bjarkason2018randomized, bjarkason2019inverse, gonzalez2018accelerating}. By contrast, data space inversion procedures \citep{sun2017new, sun2017production} essentially (approximately) marginalise over the uncertain parameters, $\bparam$, using a surrogate model, to focus on the posterior predictive distribution. This is natural when the predictions and associated uncertainties are of primary interest, as is common in geothermal settings, and avoids the need to compute the MAP estimate altogether. DSI procedures work by building an approximation to the joint distribution of the data and predictive QoIs, $\pi(\bdata, \bpred)$, using samples from this distribution generated by simulating the forward model using sets of parameters drawn from the prior. This approximation is then conditioned on the observed data, $\bdataO$, to form an approximation to the posterior predictive distribution, $\pi(\bpred|\bdataO)$. 

In the simplest form of DSI, the joint distribution of the data and predictive quantities of interest is approximated using a multivariate Gaussian distribution \citep{sun2017new}; the conditional distribution associated with a given instance of the data, $\bdataO$, is then available analytically. In many settings, however, the actual joint distribution is highly non-Gaussian, and so this approach can give poor results. In such cases, it is common to instead build a surrogate mapping that transforms samples from a simple ``reference'' distribution (typically a multivariate Gaussian distribution) to samples distributed (approximately) according to $\pi(\bdata, \bpred)$. The associated approximation to the posterior predictive distribution, under this surrogate, is then characterised using techniques such as RML \citep{jiang2020data, sun2017new, sun2017production} or ensemble methods \citep{delottier2023data, lima2020data}.

In the remainder of this section, we outline the DSI procedure we use in this work. In Section \ref{sec:map_construction}, we describe how we construct a surrogate that maps between a standard multivariate Gaussian distribution and the joint distribution of the data and predictive quantities of interest. Then, in Section \ref{sec:conditional_sampling}, we outline how we can utilise the structure of this surrogate to sample directly and efficiently from the associated approximation to the posterior predictive distribution, without the use of techniques such as RML or ensemble methods. Then, in Section \ref{sec:model_checks}, we discuss techniques for evaluating the quality of the DSI surrogate. Finally, in Section \ref{sec:related_work}, we discuss how our approach is related to other variants of the DSI algorithm, as well as other sampling techniques.

\subsubsection{Construction of the DSI Surrogate} \label{sec:map_construction}

In this section, we outline how we construct a surrogate which transforms samples distributed according to the standard multivariate Gaussian distribution of the appropriate dimension to samples distributed (approximately) according to $\pi(\bdata, \bpred)$. To do this, we first draw a set of samples, $\{\bparam^{(j)}\}_{j=1}^\numsamples$, from the prior, and a set of samples, $\{\berror^{(j)}\}_{j=1}^\numsamples$, from the distribution of the errors. We then run the forward and predictive models to generate the corresponding samples from $\pi(\bdata, \bpred)$, denoted by $\{(\bdata^{(j)}, \bpred^{(j)})\}_{j=1}^\numsamples$, where 
\begin{equation}
    \bdata^{(j)} = \bfwd(\bparam^{(j)}) + \berror^{(j)}, \qquad \bpred^{(j)} = \bpredmodel(\bparam^{(j)}).
\end{equation}
We use these samples to construct an approximate mapping between each element of the data and predictive QoIs, and a standard Gaussian random variable. To do this, we begin by noting that a mapping from a standard Gaussian random variable, $\xi^{\bdata}_{i} \sim \mathcal{N}(0, 1)$, to the $i$th element of the data, $\data_{i}$, is given by 
\begin{equation}
    \data_{i} = ((\mathcal{F}^{\bdata}_{i})^{-1} \circ \Phi)(\xi^{\bdata}_{i}), \label{eq:data_transform}
\end{equation}
where $\Phi : \mathbb{R} \rightarrow (0, 1)$ denotes the cumulative distribution function (CDF) of the standard Gaussian distribution, and $\mathcal{F}^{\bdata}_{i} : \mathcal{X}^{\bdata}_{i} \subseteq \mathbb{R} \rightarrow [0, 1]$ denotes the CDF of $\data_{i}$, which we assume is continuous and strictly increasing (and therefore invertible) on $\mathcal{X}^{\bdata}_{i}$, which denotes the range of possible values taken by $\data_{i}$. Similarly, a mapping from a standard Gaussian random variable, $\xi^{\bpred}_{i}$, to the $i$th predictive QoI, $\pred_{i}$, is given by
\begin{equation}
    \pred_{i} = ((\mathcal{F}^{\bpred}_{i})^{-1} \circ \Phi)(\xi^{\bpred}_{i}), \label{eq:pred_transform}
\end{equation}
where $\mathcal{F}^{\bpred}_{i} : \mathcal{X}^{\bpred}_{i} \subseteq \mathbb{R} \rightarrow [0, 1]$ denotes the CDF of $\pred_{i}$. We denote the mappings in Equation \eqref{eq:data_transform} and \eqref{eq:pred_transform} as $\mathcal{P}^{\bdata}_{i} := (\mathcal{F}^{\bdata}_{i})^{-1} \circ \Phi$ and $\mathcal{P}^{\bpred}_{i} := (\mathcal{F}^{\bpred}_{i})^{-1} \circ \Phi$, respectively. Then, by defining the mapping $\mathcal{P} : \mathbb{R}^{\datadim+\preddim} \rightarrow \mathbb{R}^{\datadim+\preddim}$, as
\begin{equation}
    \mathcal{P}(\bs{\xi}) := \begin{bmatrix}
        \mathcal{P}^{\bdata}(\bs{\xi}^{\bdata}) \\
        \mathcal{P}^{\bpred}(\bs{\xi}^{\bpred})
    \end{bmatrix}, \qquad \mathrm{where} \quad 
    \mathcal{P}^{\bdata}(\bs{\xi}^{\bdata}) := \begin{bmatrix}
        \mathcal{P}^{\bdata}_{1}(\xi^{\bdata}_{1}) \\
        \mathcal{P}^{\bdata}_{2}(\xi^{\bdata}_{2}) \\
        \vdots \\
        \mathcal{P}^{\bdata}_{\datadim}(\xi^{\bdata}_{\datadim})
    \end{bmatrix}, \quad
    \mathcal{P}^{\bpred}(\bs{\xi}^{\bpred}) := \begin{bmatrix}
        \mathcal{P}^{\bpred}_{1}(\xi^{\bpred}_{1}) \\
        \mathcal{P}^{\bpred}_{2}(\xi^{\bpred}_{2}) \\
        \vdots \\
        \mathcal{P}^{\bpred}_{\preddim}(\xi^{\bpred}_{\preddim})
    \end{bmatrix},
\end{equation}
it follows that 
\begin{equation}
    \begin{bmatrix} 
        \bdata \\ 
        \bpred 
    \end{bmatrix} = \mathcal{P}(\bs{\xi}), \label{eq:p_mapping}
\end{equation}
where $\bs{\xi} = [\xi_{1}^{\bdata}, \xi_{2}^{\bdata}, \dots, \xi_{\datadim}^{\bdata}, \xi_{1}^{\bpred}, \xi_{2}^{\bpred}, \dots, \xi_{\preddim}^{\bpred}]^{\top} \in \mathbb{R}^{\datadim+\preddim}$ is a vector of standard Gaussian random variables. 

\begin{figure}
    \centering
    \includegraphics[width=0.70\linewidth]{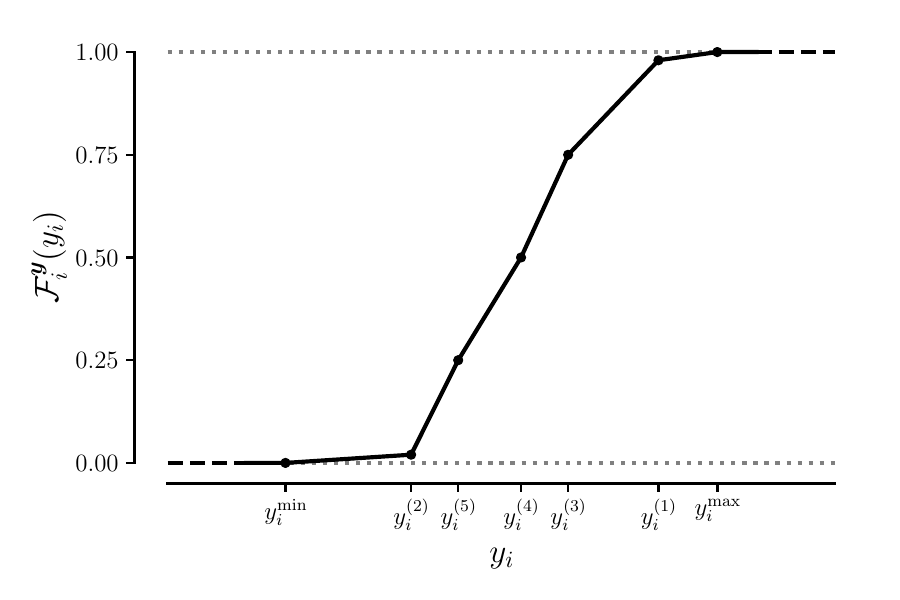}
    \caption{A piecewise linear approximation to the CDF, $\mathcal{F}^{\bdata}_{i}(\cdot)$, of a given element of the data, $\data_{i}$, using five samples, $\{\data_{i}^{(j)}\}_{j=1}^{5}$. The values $\data_{i}^{\mathrm{min}}$ and $\data_{i}^{\mathrm{max}}$ represent bounds on the range of possible values for $\data_{i}$.}
    \label{fig:piecewise_cdf}
\end{figure}

Of course, we do not have access to the CDFs $\{\mathcal{F}^{\bdata}_{i}(\cdot)\}_{i=1}^{\datadim}$ and $\{\mathcal{F}^{\bpred}_{i}(\cdot)\}_{i=1}^{\preddim}$ used to construct the mapping $\mathcal{P}$; instead, we approximate each of these functions using the previously-drawn samples. The resulting procedure is referred to as (empirical) Gaussian anamorphosis in the geostatistics literature \citep{wackernagel2003multivariate}, and is a common element in many applications of DSI \citep[see, e.g.,][]{sun2017production, sun2017new, sun2019data}. In this work, we use a piecewise linear construction of each CDF; Figure \ref{fig:piecewise_cdf} shows an example of this. Of note is the fact that the construction of the CDF shown in Figure \ref{fig:piecewise_cdf} is continuous and strictly increasing (and therefore invertible) on the interval $\mathcal{X}^{\bdata}_{i} := (\data_{i}^{\mathrm{min}}, \data_{i}^{\mathrm{max}})$, which represents the range of possible values for $\data_{i}$.

We note that, while the marginal distributions of each element of $\bs{\xi}$ are Gaussian, the joint distribution is not, in general. Nevertheless, the assumption that the joint distribution of $\bs{\xi}$ is Gaussian is often reasonable in practice; we illustrate this using the case studies in Section \ref{sec:examples}. Additionally, we note that this distribution generally possesses a non-trivial covariance structure, which we need to estimate. To do this, we first apply the inverse of transformation $\mathcal{P}$ to each of the previously-generated samples from $\pi(\bdata, \bpred)$, to obtain a set of realisations of $\bs{\xi}$; that is,
\begin{equation}
    \bs{\xi}^{(j)} = \mathcal{P}^{-1}(\bdata^{(j)}, \bpred^{(j)}), \qquad j = 1, 2, \dots, \numsamples.
\end{equation}
Next, we compute the empirical covariance matrix of these samples, which is given by
\begin{equation}
    \bs{C} = \frac{1}{\numsamples}\sum_{j=1}^{\numsamples}(\bs{\xi}^{(j)})^{\top}\bs{\xi}^{(j)}.
\end{equation}
Note that, by construction, the mean of the samples $\{\bs{\xi}^{(j)}\}_{j=1}^{\numsamples}$ is equal to $\bs{0}$.

Finally, we compute the Cholesky factorisation of $\bs{C}$, $\bs{C} = \bchol\bchol^{\top}$ \citep[see, e.g.,][]{golub2013matrix}\footnote{We note that for such a factorisation to exist, $\bs{C}$ must be full rank, which may not always be the case (note that the rank of $\bs{C}$ is always bounded above by the number of samples, $\numsamples$, used to construct it, which may be less than $\datadim+\preddim$, the dimension of $\bs{\xi}$). To avoid this, we can simply add a small (positive) multiple of the identity matrix to $\bs{C}$ to ensure it is full rank.}. Given the Cholesky factor $\bchol$, the final DSI surrogate, $\mathcal{T} : \mathbb{R}^{\datadim+\preddim} \rightarrow \mathcal{X}^{\bdata} \times \mathcal{X}^{\bpred}$, where $\mathcal{X}^{\bdata} := \mathcal{X}^{\bdata}_{1} \times \mathcal{X}^{\bdata}_{2} \times \cdots \times \mathcal{X}^{\bdata}_{\datadim} \subseteq \mathbb{R}^{\datadim}$ and $\mathcal{X}^{\bpred} := \mathcal{X}^{\bpred}_{1} \times \mathcal{X}^{\bpred}_{2} \times \cdots \times \mathcal{X}^{\bpred}_{\preddim} \subseteq \mathbb{R}^{\preddim}$, which transforms samples of the Gaussian random variable $\bs{\eta} \sim \mathcal{N}(\bs{0}_{\datadim+\preddim}, \bs{I}_{\datadim+\preddim})$ to samples distributed approximately according to $\pi(\bdata, \bpred)$, can be expressed as 

\begin{equation}
    \mathcal{T}(\bs{\eta}) = (\mathcal{P} \circ \bchol)(\bs{\eta}). \label{eq:dsi_mapping}
\end{equation}

Note that this transformation is composed of two key steps. First, the application of $\bchol$ to $\bs{\eta}$ acts to give $\bs{\eta}$ the required correlation structure. Next, the application of $\mathcal{P}$ acts to transform the marginal distributions of each element of the resulting vector to the desired form.

\subsubsection{Posterior Predictive Simulation} \label{sec:conditional_sampling}

Once the mapping $\mathcal{T}(\cdot)$ has been constructed, it is easy to draw samples from the approximation, under $\mathcal{T}(\cdot)$, to the posterior predictive distribution, $\pi(\bpred|\bdataO)$. To illustrate how this can be done, we first note that the mapping is endowed with a particular structure. Because the Cholesky factor $\bchol$ is a lower triangular matrix, it follows that $\mathcal{T}(\cdot)$ is a lower triangular transformation; that is, each component $\mathcal{T}_{i}(\cdot)$ is a function of the first $i$ input variables only. Additionally, each component $\mathcal{T}_{i}(\cdot)$ is strictly increasing as a function of the $i$th input variable; this is a consequence of the fact that diagonal entry $\chol_{ii}$ of the Cholesky factor is strictly positive by definition, and the mapping $\mathcal{P}_{i}(\cdot)$ is strictly increasing by construction.

The triangular structure of $\mathcal{T}(\cdot)$ means that it can be expressed in the form
\begin{equation}
    \mathcal{T}(\bs{\eta}) = \begin{bmatrix}
        \mathcal{T}^{\bdata}(\bs{\eta}_{\bdata}) \\
        \mathcal{T}^{\bpred}(\bs{\eta}_{\bdata}, \bs{\eta}_{\bpred})
    \end{bmatrix},
\end{equation}
where $(\bs{\eta}_{\bdata}, \bs{\eta}_{\bpred})$ is a partitioning of the random variable $\bs{\eta}$ into its first $d$ components and its last $m$ components, and $\mathcal{T}^{\bdata} : \mathbb{R}^{\datadim} \rightarrow \mathcal{X}^{\bdata}$ and $\mathcal{T}^{\bpred} : \mathbb{R}^{\datadim+\preddim} \rightarrow \mathcal{X}^{\bpred}$. Then, owing to the monotonicity property of $\mathcal{T}$, it follows that the mapping 
\begin{equation}
    \bs{\eta}_{\bpred} \mapsto \mathcal{T}^{\bpred}(\bs{\eta}_{\bdata}^{\mathrm{obs}}, \bs{\eta}_{\bpred}),
\end{equation}
where 
\begin{equation}
    \bs{\eta}_{\bdata}^{\mathrm{obs}} = (\mathcal{T}^{\bdata})^{-1}(\bdata^{\mathrm{obs}}),
\end{equation}
transforms a vector, $\bs{\eta}_{\bpred}$, of independent samples from the standard normal distribution to samples distributed according to the approximation, under $\mathcal{T}(\cdot)$, of the conditional density of $\bpred\,|\,\bdata^{\mathrm{obs}}$; that is, the posterior predictive distribution $\pi(\bpred|\bs{\bdataO})$. For further details and a proof of this property, we refer the reader to \citet[][Sec.~7]{marzouk2016sampling}. 

The above gives us a simple and efficient way to generate samples from the posterior predictive distribution. We first apply the inverse of $\mathcal{T}^{\bdata}$ to the observations, $\bdata^{\mathrm{obs}}$, to obtain $\bs{\eta}_{\bdata}^{\mathrm{obs}}$. We then generate a sample by computing 
\begin{equation}
    \mathcal{T}^{\bpred}(\bs{\eta}_{\bdata}^{\mathrm{obs}}, \bs{\eta}_{\bpred}), \qquad \mathrm{where} \quad \bs{\eta}_{\bpred} \sim \mathcal{N}(\bs{0}_{\preddim}, \bs{I}_{\preddim}).
\end{equation}

\subsubsection{Evaluating the Quality of the DSI Surrogate} \label{sec:model_checks}

The construction of the DSI surrogate outlined in Section \ref{sec:map_construction} makes use of two approximations; the quality of the DSI approximation to the posterior predictive distribution depends on the degree to which these approximations are accurate. The first is the approximation of the CDF of each element of the data and predictive quantities of interest using the set of samples drawn from the prior. This approximation will, in general, improve as the number of samples, $\numsamples$, used to construct each CDF is increased. The second is the approximation of the joint density of the random variable $\bs{\xi}$, obtained by applying the inverse of the mapping $\mathcal{P}(\cdot)$ to the data and predictive quantities of interest (see Eq.~\ref{eq:p_mapping}) as Gaussian. 

There are several qualitative checks we can carry out to evaluate the quality of the DSI surrogate, $\mathcal{T}(\cdot)$, after its construction. First, we can plot the joint densities of elements of the samples of the data and predictive quantities of interest after applying the inverse of the surrogate mapping to each sample, to evaluate their degree of Gaussianity. Ideally, the resulting samples should appear similar to draws from a standard multivariate Gaussian distribution. However, the data and predictive quantities of interest are generally high-dimensional, which makes this check challenging to carry out. A second approach is to plot samples from the approximation to the distribution of the data and/or predictive quantities of interest under the DSI surrogate, by applying the mapping to sets of independent samples from the standard normal distribution \citep[see, e.g.,][]{sun2017production, jiang2021data}. Ideally, the resulting samples should appear qualitatively similar to the samples used to construct the surrogate. Finally, we can withhold a set of ``validation'' samples of the data and predictive quantities of interest when constructing the DSI surrogate. We can then condition on each validation dataset using DSI to form the associated approximation to the posterior predictive distribution, and determine whether the corresponding values of the predictive quantities of interest fall within the range of the posterior predictions. We note that the construction of the DSI surrogate (including the required simulations of the forward model) can be done \emph{offline}; that is, without knowledge of the data. As a result, this process can be carried out in a computationally efficient manner. We use the case studies in Section \ref{sec:examples} to illustrate the second and third approaches to evaluating the quality of the DSI surrogate.

Additionally, we note that, like many other inference techniques that use surrogate models, DSI may not provide accurate results when used to condition on observed data that differs meaningfully to the samples of the data used to build the DSI surrogate. To deal with this, we can make use of prior predictive checks \citep[see, e.g,][]{gelman2013bayesian, gelman2020bayesianworkflow}---that is, visualisations of samples of the data (and predictive quantities of interest, if the data comprise a subset of these) drawn from the prior---to ensure that these samples encompass the full range of expected behaviour of the system under consideration. If this is not the case, the prior can be re-characterised and the process repeated. This will help to reduce the risk that the prior is unrepresentative of the data that is collected. We note that similar ideas are also part of the Bayesian evidential learning framework \citep{scheidt2018quantifying}.

Finally, we emphasise that the DSI surrogate may generate samples that do not adhere to the physics of the particular problem under consideration. It is, therefore, important that results obtained using the DSI algorithm (or any other form of surrogate model) are interpreted with caution, by people with a high level of domain knowledge. If a significant number of the samples generated using DSI do not appear to be physically plausible, this suggests that an alternative approach to inference may be required.

\subsubsection{Related Work} \label{sec:related_work}

We note that the DSI algorithm we have described shares similarities to the variant of DSI introduced by \citet[][Sec.~3]{sun2017new}, which we briefly outline. In this variant of DSI, the authors first perform principal component analysis (PCA) on a set of samples of the data and predictive quantities of interest obtained by simulating the forward model using a set of samples drawn from the prior. Then, sampling from the resulting approximation to the prior predictive distribution of the data and predictive quantities of interest involves the projection of samples from the standard multivariate Gaussian distribution of the appropriate dimension onto the previously-computed principal components, followed by an empirical Gaussian anamorphosis step which acts to transform the marginal distributions of each quantity to the desired shape. We note that the use of PCA to re-parametrise the observations and predictive quantities of interest means that the resulting mapping is not lower triangular, making it challenging to sample from directly; instead, \citet{sun2017new} characterise the approximation to the posterior predictive distribution associated with a particular instance of the data using a randomised maximum likelihood (RML) procedure. Our variant of the DSI algorithm, by contrast, allows for direct simulation from the posterior predictive distribution, avoiding the need for sampling methods such as RML, which add additional computation and an extra degree of approximation to the procedure.

Additionally, we note that the DSI mapping described in this work can be thought of as a simple instance of a larger class of lower triangular, strictly increasing transformations that provide an (approximate) mapping between a simple ``reference'' random variable and a complex ``target'' random variable. For further discussion of these ideas, we refer the reader to \citet{marzouk2016sampling}.

\section{Computational Examples}\label{sec:examples}

We now demonstrate the DSI approach by applying it to a simplified two-dimensional problem, as well as a large-scale, three-dimensional reservoir model. Both problems are adapted from those presented in \citet{de2024ensemble}. We first outline the governing equations for general geothermal reservoir modelling, before presenting the results of each model problem.

\subsection{Governing Equations}\label{sec:equations}

The dynamics of a geothermal reservoir are described (mathematically) by a non-isothermal, multi-phase version of Darcy's law, enforcing conservation of mass and energy \citep{croucher2020waiwera, osullivan2016reservoir}. In what follows, we denote by $\Omega\in\mathbb{R}^3$ the domain of interest, with boundary $\partial\Omega$ and outward-facing normal vector $\bs{n}$. Furthermore, we let $\CompNum$ denote the number of each component (e.g., water, air, energy). The governing equations, expressed in integral form, are then
\begin{subequations}\label{eq:forward}
\begin{align} \label{eq:balances}
    \frac{\dd}{\dd t}\int_\Omega M^{\CompNum}\,\dd\bs{x} = -\int_{\partial\Omega}\bs{F}^{\CompNum}\cdot\bs{n}\,\dd\bs{\sigma} +\int_\Omega q^{\CompNum}\,\dd\bs{x},\quad \CompNum = 1, 2, \dots, N+1,
\end{align}
where components $\CompNum = 1, 2, \dots, N$ denote the mass components and component $\CompNum = N+1$ denotes the energy component. For each mass component, $M^{\CompNum}$ denotes mass density (kg\,m$^{-3}$), $\bs{F}^{\CompNum}$ denotes mass flux (kg\,m$^{-2}$\,s$^{-1}$), and $q^{\CompNum}$ denotes mass sources or sinks (kg\,m$^{-3}$\,s$^{-1}$). For the energy component, $M^{\CompNum}$ denotes energy density (J\,m$^{-3}$), $\bs{F}^{\CompNum}$ denotes energy flux (J\,m$^{-2}$\,s$^{-1}$), and $q^{\CompNum}$ denotes energy sources or sinks (J\,m$^{-3}$\,s$^{-1}$). The mass and energy densities can be expressed as 
\begin{align}
    M^{\CompNum} = \begin{cases}
        \phi(\rho_{\rm l}S_{\rm l}X_{\rm l}^{\CompNum} + \rho_{\rm v}S_{\rm v}X_{\rm v}^{\CompNum}), & \CompNum< N+1,\\
        (1-\phi)\rho_{\rm r}u_{\rm r}T+\phi(\rho_{\rm l}u_{\rm l}S_{\rm l}+\rho_{\rm v}u_{\rm v}S_{\rm v}), & \CompNum=N+1,
    \end{cases}
\end{align}
where $\phi$ denotes porosity (dimensionless), $S_{\rm l}$ and $S_{\rm v}$ denote liquid saturation and vapour saturation (dimensionless) respectively, $\rho_{\rm l}$, $\rho_{\rm v}$ and $\rho_{\rm r}$ denote the density of the liquid, vapour and rock (kg\,m$^{-3}$) respectively, $X_{\rm l}^{\CompNum}$ and $X_{\rm v}^{\CompNum}$ denote the liquid and vapour mass fractions (dimensionless) of component $\CompNum$ respectively, $u_{\rm l}$ and $u_{\rm v}$ denote the internal energy of the liquid and vapour (J\,kg$^{-1}$) respectively, $u_{\rm r}$ denotes the specific heat of the rock (J\,kg$^{-1}$\,K$^{-1}$), and $T$ denotes temperature (K). Next, the mass ($\CompNum<N+1$) fluxes are given by the sum of the mass flux of liquid and the mass flux of vapour,
\begin{align}
    \bs{F}^{\CompNum} = \bs{F}^{\CompNum}_{\rm l}+\bs{F}^{\CompNum}_{\rm v},\quad
    \bs{F}^{\CompNum}_{\rm l} = -\frac{\bs{k}k_{\rm rl}}{\nu_{\rm l}}X_{\rm l}^{\CompNum}(\nabla \Pressure-\rho_{\rm l}\bs{g}),\quad
    \bs{F}^{\CompNum}_{\rm v} = -\frac{\bs{k}k_{\rm rv}}{\nu_{\rm v}}X_{\rm v}^{\CompNum}(\nabla \Pressure-\rho_{\rm v}\bs{g}).
\end{align}
Here, $\bs{\Perm}$ represents the permeability tensor (m$^2$), $\Pressure$ denotes pressure (Pa), $\nu_{\rm l}$ and $\nu_{\rm v}$ denote the kinematic viscosity of liquid and vapour (m$^2$\,s$^{-1}$) respectively, $\Perm_{\rm rl}$ and $\Perm_{\rm rv}$ denote relative permeabilities (dimensionless), and $\bs{g}$ denotes gravitational acceleration (m\,s$^{-2}$). Finally, the energy ($\CompNum=N+1$) flux is given by 
\begin{align}
    \bs{F}^{\CompNum}=-K\nabla T+\sum_{m=1}^{N}\sum_\chi h^m_\chi 
    \bs{F}^m_\chi,
\end{align}
\end{subequations}
where $h^m_\chi$ denotes the specific enthalpy (J\,kg$^{-1}$) of mass component $m$ in phase $\chi$, and $K$ denotes thermal conductivity (J\,s$^{-1}$\,m$^{-1}$\,K$^{-1}$). 

\subsection{Two-Dimensional Single-Phase Model} \label{sec:Darcy}

Our first model problem serves to provide a comparison between the posterior predictions produced using MCMC, linearisation about the MAP estimate (see Section \ref{sec:local_lin}), and DSI, when applied to a high-dimensional subsurface flow problem. Problems of a similar nature are often used as benchmarks when evaluating uncertainty quantification algorithms \citep[see, e.g.,][]{aristoff2023benchmark, christie2001tenth}.

\subsubsection{Problem Setup}

For this problem, we make several simplifications to the governing equations introduced in Section \ref{sec:equations}. Namely, we consider a two-dimensional reservoir with domain $\Omega = (0\,\rm{\preddim}, 1000\,\rm{\preddim})^{2}$, containing single-phase, isothermal, slightly compressible fluid \citep[see, e.g.,][]{chen2007reservoir}, with a set of $n_{w}$ production wells. We further assume that the permeability tensor, $\bs{\Perm}$, is isotropic. In this case, Equation \eqref{eq:forward} simplifies to
\begin{equation}
    c\phi\frac{\partial \Pressure}{\partial t} - \frac{1}{\mu}\nabla\cdot(\Perm\nabla \Pressure) = \sum_{i=1}^{n_{w}}q_{i}\delta(\x-\x_{i}), \qquad \x \in \Omega, \,\, t \in (0, \tau],
\end{equation}
where $c$ denotes fluid compressibility (assumed to be $2.9\times10^{-8}\,{\rm Pa}^{-1}$), $\mu$ denotes dynamic viscosity (assumed to be $0.5\,{\rm mPa}\,{\rm s}$), $q_{i}$ denotes the extraction rate of well $i$, and $\delta(\x-\x_{i})$ denotes a Dirac delta mass centred at well $i$, the location of which is indicated by $\x_{i}$. We assume a reservoir porosity of $\phi=0.3$, and impose the boundary and initial conditions
\begin{equation}
\begin{aligned}
    -\Perm\nabla \Pressure\cdot\bs{n} &= 0, \qquad && \x \in \partial\Omega, \,\, &&t \in (0, \tau], \\
    \Pressure &= \Pressure_{0}, \qquad && \x \in \Omega, \,\, &&t = 0. \label{eq:darcy_bcs}
\end{aligned}
\end{equation}
In the above, $\Pressure_{0}$ denotes the initial reservoir pressure, which we assume to be $20\,{\rm MPa}$. The sole uncertain parameter is the (spatially varying, isotropic) permeability of the reservoir, $\Perm$.

\begin{figure}
    \centering
    \includegraphics[width=0.95\textwidth]{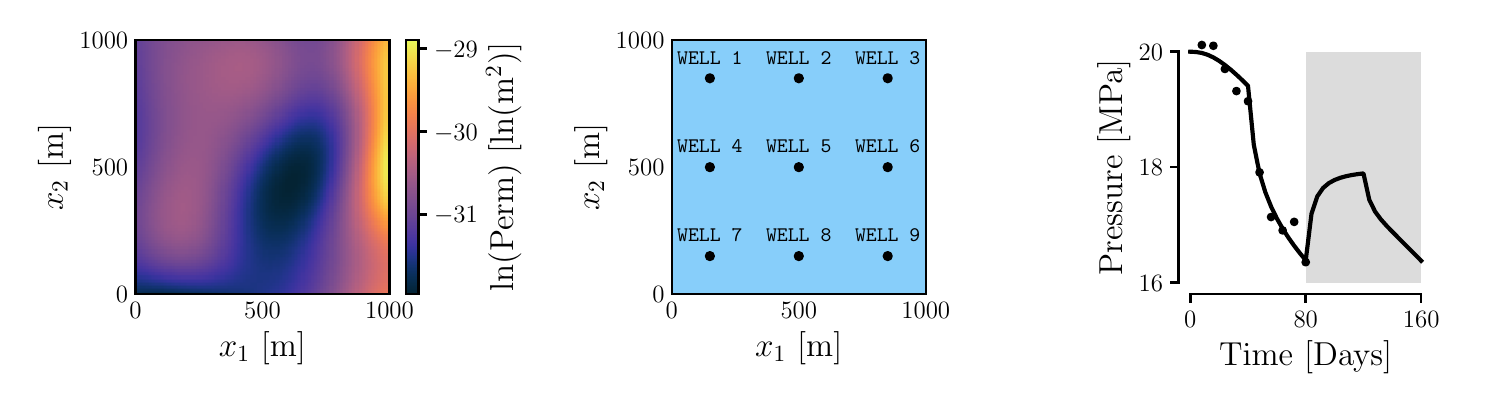}
    \caption{The setup for the simplified two-dimensional reservoir model. Left: the true log-permeability field. Centre: the locations of the production wells. Right: the pressure in well 8; the solid line denotes the true pressure, the dots denote the noisy data collected during the production history period, and the grey region denotes the forecast period.}
    \label{fig:darcy_setup}
\end{figure}

\begin{figure}
    \centering
    \includegraphics[width=0.95\textwidth]{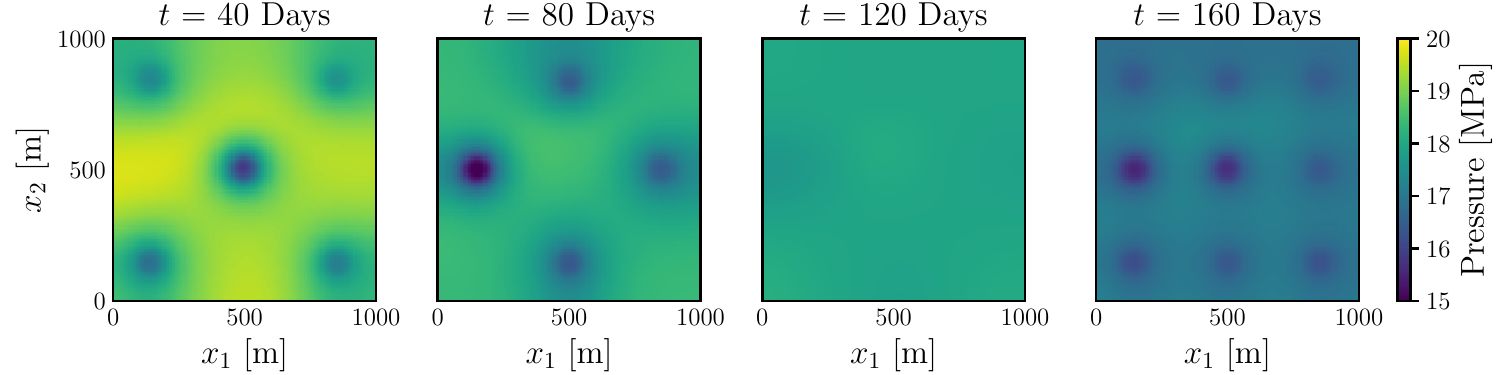}
    \caption{The true reservoir pressure at $t=40$, $t=80$, $t=120$ and $t=160$ days.}
    \label{fig:darcy_pressure}
\end{figure}

We consider a setup in which a set of $n_{w} = 9$ production wells operate over a period of $\tau = 160$ days. The first $80$ days act as the production period, while the final $80$ days act as the forecast period. The location of each well is indicated in Figure \ref{fig:darcy_setup}. For the first $40$ days, each of the odd-numbered wells extracts fluid at a rate of $50\,{\rm m}^{3}\,{\rm day}^{-1}$, while the even-numbered wells are turned off. For the next $40$ days, this is reversed; the even-numbered wells extract fluid at a rate of $50\,{\rm m}^{3}\,{\rm day}^{-1}$, while the odd-numbered wells are turned off. For the next 40 days, all wells are turned off, before operating at a rate of $25\,{\rm m}^{3}\,{\rm day}^{-1}$ for the final 40 days. Figure \ref{fig:darcy_pressure} shows the true reservoir pressure at the end of each 40-day period.

\subsubsection{Prior Parametrisation} \label{sec:darcy_params}

\begin{figure}
    \centering
    \includegraphics[width=0.95\textwidth]{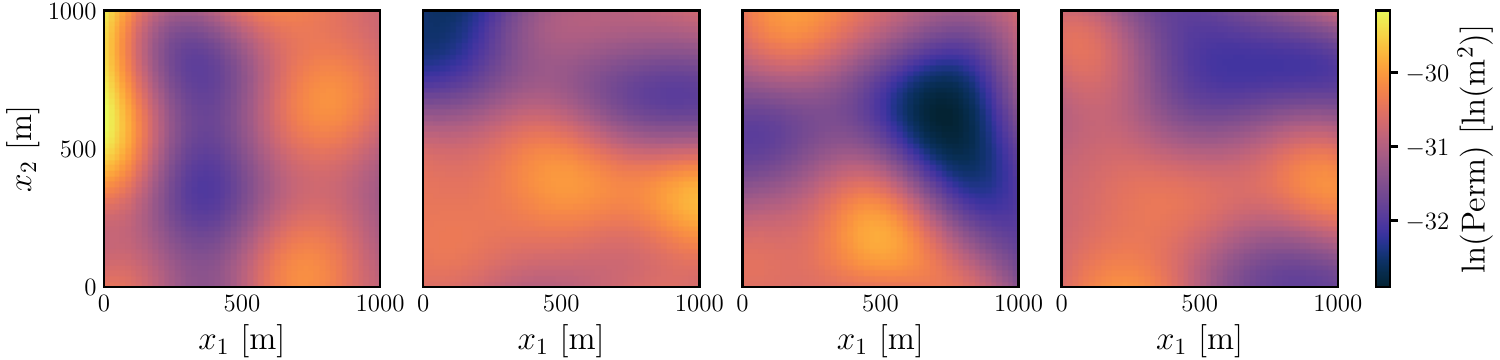}
    \caption{Samples from the prior distribution of the reservoir permeability.}
    \label{fig:prior_samples}
\end{figure}

As is standard, when solving the inverse problem we work in terms of the log-permeability, $u := \ln(\Perm)$, which ensures that the resulting estimates of the permeability are positive. We parametrise the log-permeability of the reservoir using a Gaussian random field (GRF) with a mean function of $m(\bs{x}) = -31\,\ln({\rm m}^{2})$, and a squared-exponential covariance function \citep{williams2006gaussian}, given by
\begin{equation}
    \mathcal{C}(\x, \x') = \sigma^{2}\exp\left(-\frac{1}{2\lengthscale^{2}}\norm{\x-\x'}^{2}\right). \label{eq:sq_exp}
\end{equation}
We use a standard deviation of $\sigma=0.75\,\ln({\rm m}^{2})$ and a characteristic lengthscale of $\lengthscale=250\,{\rm m}$. To reduce the dimension of the parameter space and accelerate the convergence of our MCMC sampler, we approximate this GRF using a truncated Karhunen--Lo{\`e}ve expansion; that is,
\begin{equation}
    \bs{u} \approx \sum_{i=1}^{n}\sqrt{\lambda_{i}}\bs{v}_{i}\eta_{i}, \label{eq:kl}
\end{equation}
where $\eta_{i} \sim \mathcal{N}(0, 1)$. In Equation \eqref{eq:kl}, $\{(\lambda_{i}, \bs{v}_{i})\}^{n}_{i=1}$ denote the $n$ largest eigenpairs of the covariance matrix of the (discretised) GRF, where $n$ is typically small compared to the dimension of the field. Under this parametrisation, the values of the coefficients $\{\xi_{i}\}^{n}_{i=1}$ become the targets of inference. Here, we retain $n=50$ coefficients. Figure \ref{fig:prior_samples} shows several draws from the prior.

\subsubsection{Data} 

We assume that the pressure at each well is recorded every 8 days throughout the production history period (i.e., the first $80$ days); this gives a total of $90$ measurements. We add independent Gaussian noise with a standard deviation of $1\%$ of the initial reservoir pressure to each observation. Figure \ref{fig:darcy_setup} shows the data collected at well 8.

\subsubsection{Simulation}

We discretise the system using a cell-centred finite difference scheme \citep{chen2007reservoir, haber2007model}, and use the backward Euler method to solve for the dynamics of the system over time. The permeability distribution of the true system, shown in Figure \ref{fig:darcy_setup}, is generated using a draw from the prior. To avoid the ``inverse crime'' of generating the synthetic data and solving the inverse problem using the same numerical discretisation \citep{kaipio2006statistical, kaipio2007statistical}, we use an $80\times80$ grid when simulating the dynamics of the true system, but a $50\times50$ grid when carrying out each inversion.

\subsubsection{Inference Methods}

We aim to use the data collected at each well over the production period to estimate how the pressure at each well will change over the forecast period.

We compute a complete characterisation of the posterior and posterior predictive distributions using the preconditioned Crank-Nicolson MCMC sampler \citep{chen2019dimensionrobust, cotter2013mcmc}, which is commonly used to solve high-dimensional inverse problems. We run four Markov chains, each initialised at a random draw from the prior, for 500,000 iterations, and discard the first half of each chain as burn in. These results provide a reference to which we can compare the posterior predictive distributions produced using linearisation about the MAP estimate and DSI. 

When characterising the posterior using LMAP (as outlined in Section \ref{sec:local_lin}), we compute the MAP estimate using a matrix-free inexact Gauss-Newton conjugate gradient method \citep[see, e.g.,][]{haber2007model, petra2011model}. The process of computing the MAP estimate and forming the approximate posterior covariance matrix requires 244 ``forward-like'' solves (these include both forward and adjoint solves, which are associated with similar computational costs), though this could, of course, be reduced through the use of an improved optimisation method. We then run samples from the approximate posterior through the predictive model to obtain samples from the corresponding (approximate) posterior predictive distribution. Note that this corresponds to the first method of approximate predictive uncertainty quantification outlined in Section \ref{sec:pred_uq_approx}.

When approximating the posterior predictive distribution using DSI, we use an initial set of 1000 samples from the prior to build the mapping outlined in Section \ref{sec:map_construction}. We then examine the differences in the results when the number of samples used to estimate the DSI mapping is varied.

\subsubsection{Validation}

\begin{figure}
    \centering
    \includegraphics[width=0.95\linewidth]{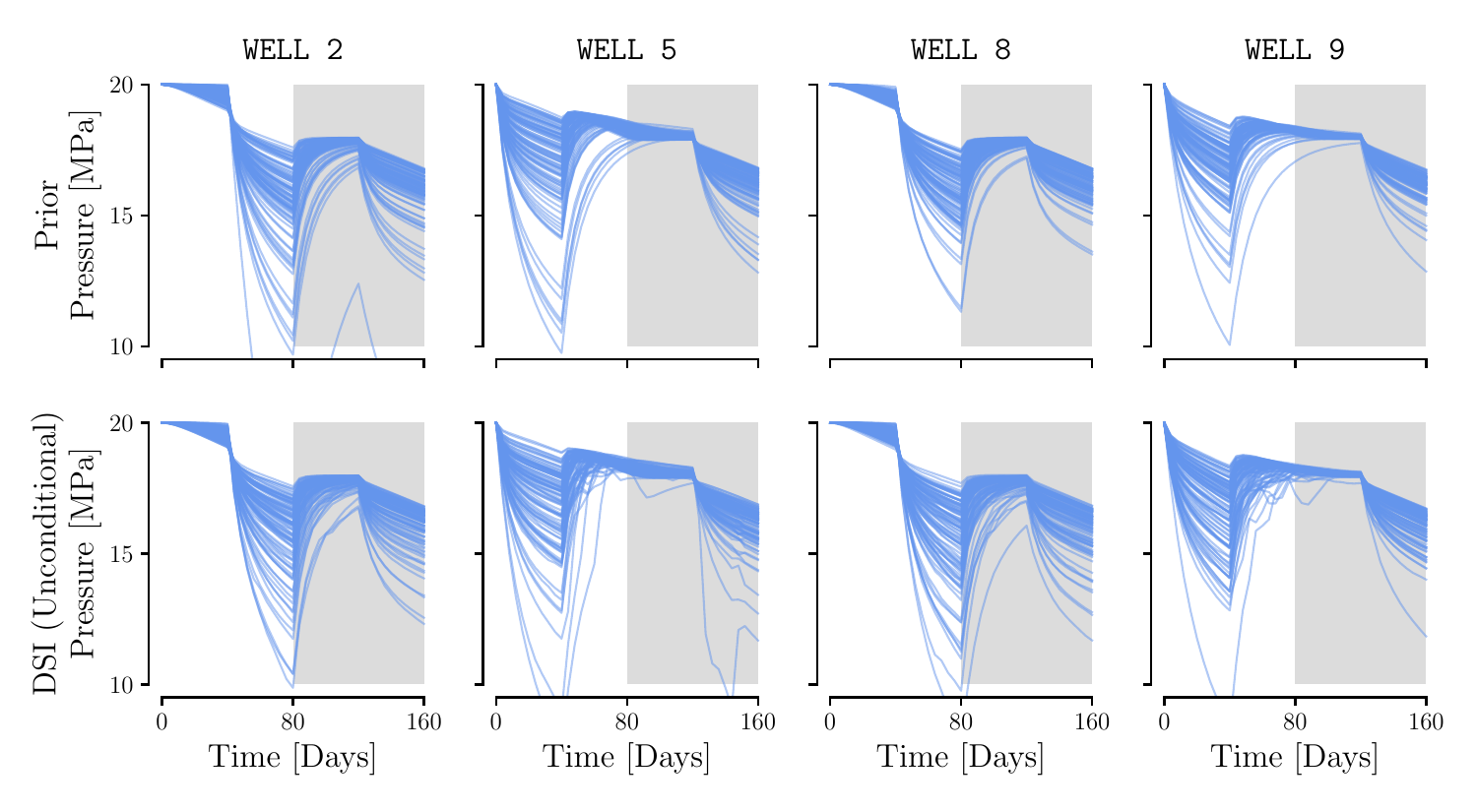}
    \caption{Sets of 100 samples from the prior predictive distribution (top row) and the DSI approximation to the prior predictive distribution (bottom row) of the pressure in wells 2, 5, 8 and 9. In all plots, the grey region denotes the forecast period.}
    \label{fig:prior_vs_dsi}
\end{figure}

Before discussing the results obtained using each inference technique, we evaluate the quality of the DSI surrogate, $\mathcal{T}(\cdot)$, by plotting a set of (unconditional) realisations from the DSI approximation to the prior predictive distribution of the pressure of several of the wells of the system, as discussed in Section \ref{sec:model_checks}. These results are shown in Figure \ref{fig:prior_vs_dsi}; we note that this visualisation also functions as an example of a prior predictive check. We observe that in general, the DSI predictions exhibit similar behaviour to the prior predictions, with the exception of a handful of samples of the pressure in wells 5 and 9 which exhibit some oscillatory behaviour near the end of the production period that is not present in the corresponding samples from the prior.


\subsubsection{Results}

\begin{figure}
    \centering
    \includegraphics[width=0.95\textwidth]{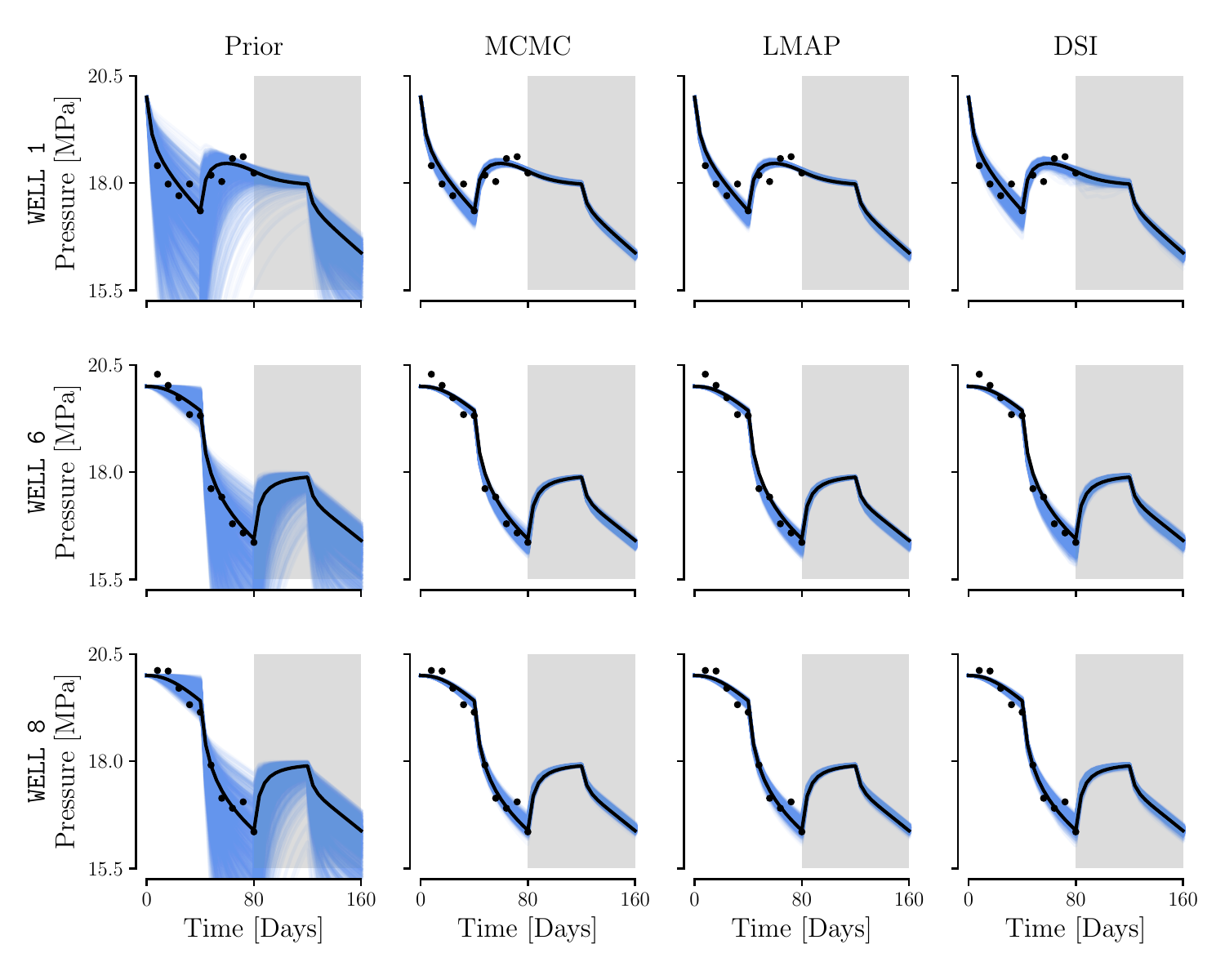}
    \caption{Sets of $1000$ samples from the prior predictive distribution (left) and the posterior predictive distributions generated using MCMC (centre left), LMAP (centre right), and DSI (right), for wells 1, 6 and 8. In all plots, the blue lines indicate the samples, the black line denotes the true well pressure, the black dots denote the observations collected during the production history period, and the grey region denotes the forecast period.}
    \label{fig:darcy_post_samples}
\end{figure}

\begin{figure}
    \centering
    \includegraphics[width=0.75\textwidth]{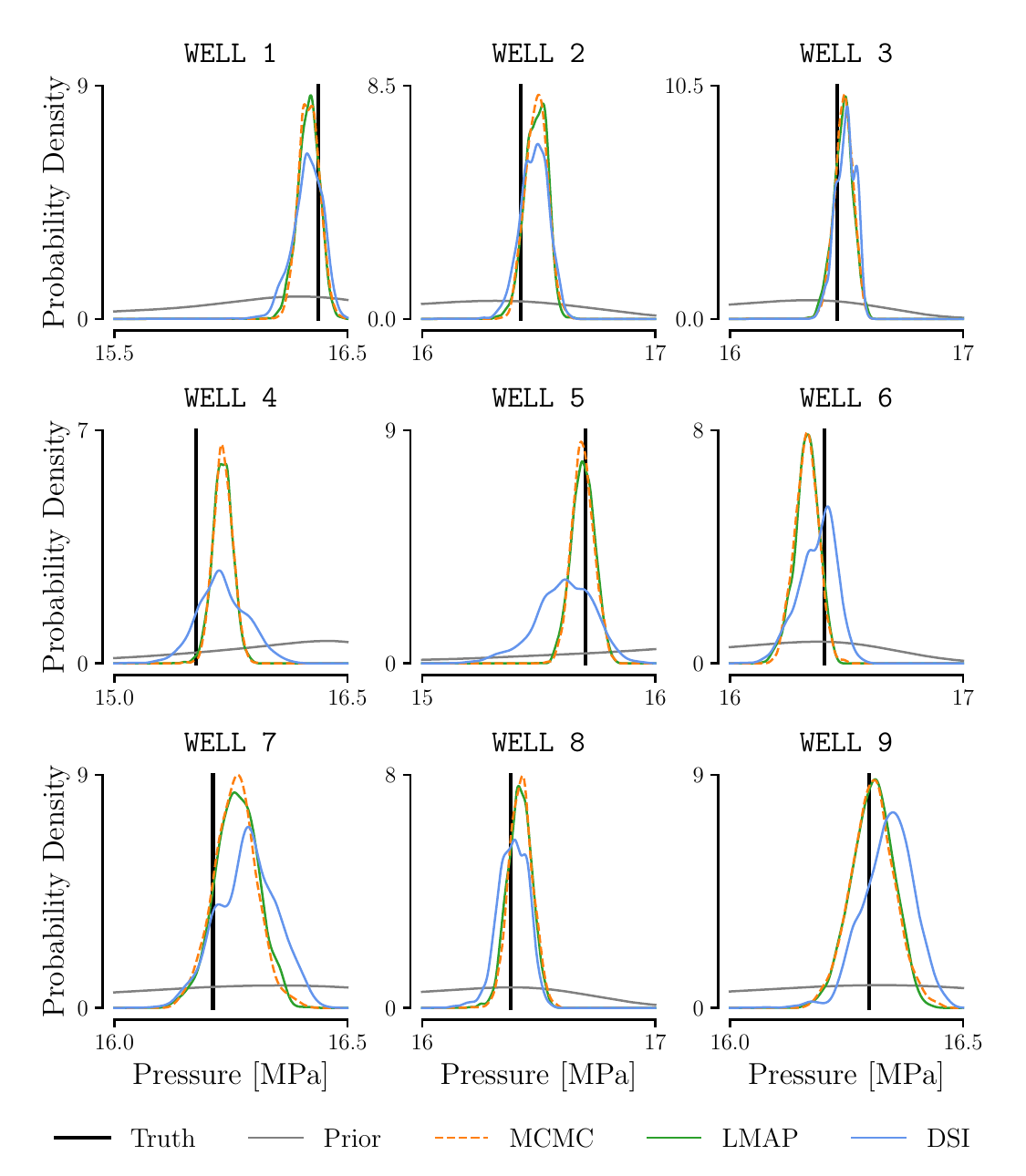}
    \caption{The prior predictive distribution and posterior predictive distributions of the pressure in each well at the end of the forecast period ($t = 160$ days) obtained using MCMC, LMAP, and DSI.}
    \label{fig:darcy_final_pressures}
\end{figure}

We now compare the results obtained using each inference technique. Figure \ref{fig:darcy_post_samples} shows a set of 1000 samples of the pressure at wells 1, 6 and 8, drawn from the posterior predictive distributions produced using MCMC, LMAP, and DSI. In each case, a set of 1000 samples from the prior is also presented for comparison. We observe that in all cases, the posterior uncertainty is significantly reduced in comparison to the prior uncertainty, and the true pressure at each well is contained within the predictions. The predictions of the pressure at well 1 obtained using DSI appear to have a slightly greater variance in comparison to those generated using MCMC and LMAP. Those for well 6 and well 8, by contrast, appear very similar. Figure \ref{fig:darcy_final_pressures} shows the estimates of the marginal densities of the pressure in each well at the end of the forecast period ($t = 160$ days) obtained using each method (we note that the DSI densities are known analytically). These plots largely reinforce these conclusions.

\begin{figure}
    \centering
    \includegraphics[width=0.75\textwidth]{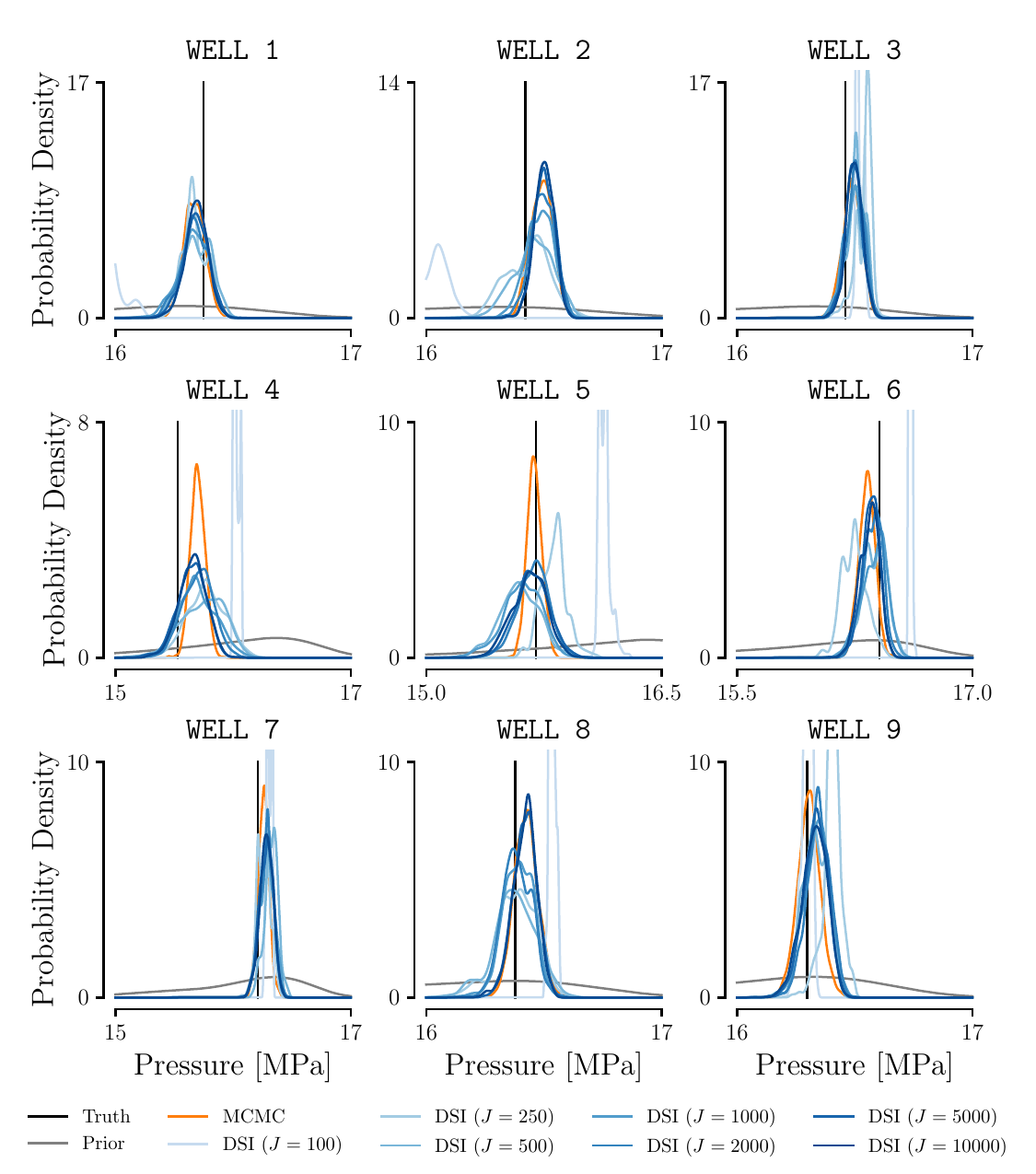}
    \caption{The prior predictive distribution and posterior predictive distributions of the pressure in each well at the end of the production period ($t = 160$ days) obtained using MCMC, and DSI with varying numbers of samples.}
    \label{fig:sample_comp}
\end{figure}

Figure \ref{fig:sample_comp} shows how the posterior predictions change as the number of samples used to estimate the DSI surrogate $\mathcal{T}(\cdot)$ varies. The posterior predictive distribution generated using $100$ samples is often significantly different to the approximations generated using larger numbers of samples (see, e.g., well 2 and well 3). However, after the number of samples reaches $500$, the predictive distributions begin to look very similar to one another. This suggests that $\numsamples = 500$ or more samples is an appropriate number to use when applying DSI to this problem. We note, however, that there is no guarantee a sample size that provides acceptable results in one context will provide acceptable results in another. The number of samples after which the DSI estimate of the predictive QoIs begins to stabilise will depend on a variety of factors, including the characteristics of the forward model, the prior parametrisation, and the dimensions of the predictive quantities of interest and the data. In future work, it would be valuable to investigate how the required number of samples for a stable DSI estimate of the predictive QoIs varies when these characteristics of the problem are modified.

\subsection{Three-Dimensional Reservoir Model} \label{sec:geo}

The second test case we consider is a synthetic three-dimensional reservoir model.

\subsubsection{Problem Setup}

\begin{figure}
    \centering
    \includegraphics[width=1.0\textwidth]{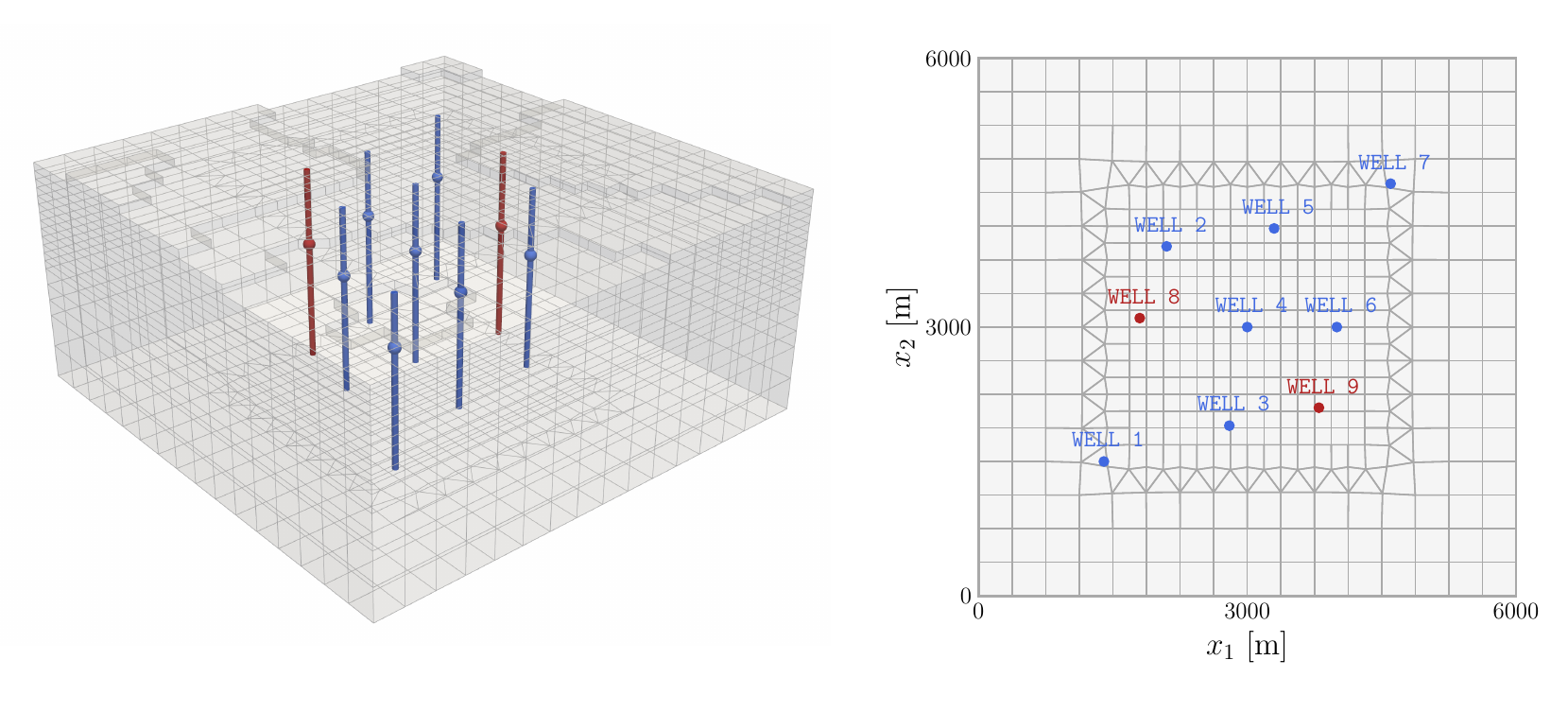}
    \caption{The mesh of the synthetic reservoir model, and the locations of each production well. Existing wells are denoted using blue and new wells are denoted using red.}
    \label{fig:geo_mesh}
\end{figure}

\begin{figure}
    \centering
    \includegraphics[width=1.0\textwidth]{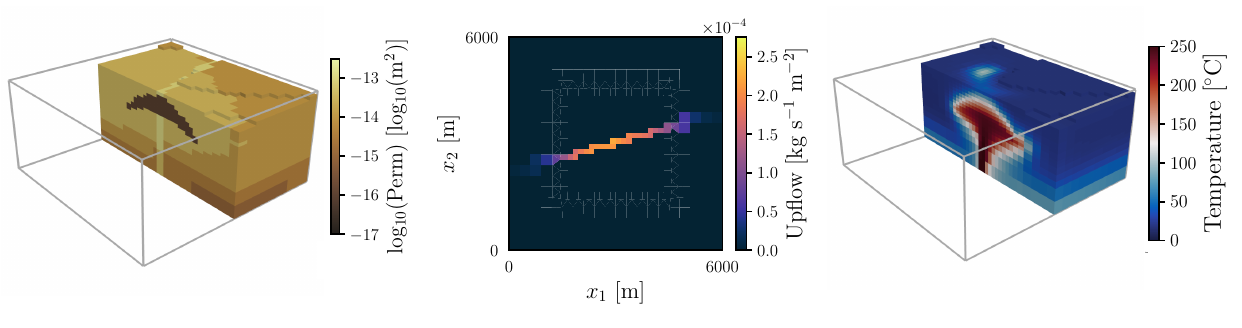}
    \caption{The true permeability structure (left), mass upflows (centre) and natural state convective plume (right) of the synthetic reservoir model.}
    \label{fig:geo_truth}
\end{figure}

The model domain, shown in Figure \ref{fig:geo_mesh}, spans 6000\,m in the horizontal ($x_{1}$ and $x_{2}$) directions, and extends to a depth of 3000\,m in the vertical direction. Figure \ref{fig:geo_truth} shows the true subsurface permeability structure, mass upflows, and natural state convective plume of the system.

We assume that we have been extracting fluid at each of the seven existing wells (wells 1--7 in Figure \ref{fig:geo_mesh}) of the system at a rate of 0.25\,kg\,s$^{-1}$ over a production history period of one year. We then wish to estimate the downhole temperature profiles associated with each of the existing wells, as well as two new wells (well 8 and well 9 in Figure \ref{fig:geo_mesh}), and to predict how the pressure and enthalpy of the fluid extracted at each well will change if we operate all wells at the increased rate of 0.5\,kg\,s$^{-1}$ for a forecast period of an additional year. Each well has a single feedzone at a depth of 1200\,m. As is standard in geothermal reservoir modelling, we consider a combined natural state and production history simulation \citep{osullivan2016reservoir}; that is, we simulate the dynamics of the system until steady-state conditions are reached, then use the resulting state of the system as the initial condition for the subsequent production simulation.

\subsubsection{Prior Parametrisation}

When parametrising the prior, we consider uncertainty in the subsurface permeability structure (modelled as isotropic, for simplicity), the reservoir porosity, and the location and magnitude of the hot mass upflow at the base of the reservoir. All other reservoir properties are assumed known. The rock of the reservoir is assumed to have a thermal conductivity of 2.5\,W\,m$^{-1}$\,K$^{-1}$ and a specific heat of 1000\,J\,kg$^{-1}$\,K$^{-1}$. The top boundary of the model is set to a constant pressure of 1\,bar and a temperature of 20$^{\circ}$C; this represents an atmospheric boundary condition. The side boundaries are closed. We impose a background heat flux of 200\,mW\,m$^{-2}$ through all cells on the bottom boundary except for those which are associated with a mass upflow. All of the mass upflow entering through the bottom boundary is associated with an enthalpy of 1500\,kJ\,kg$^{-1}$.

\begin{figure}
    \centering
    \includegraphics[width=0.95\textwidth]{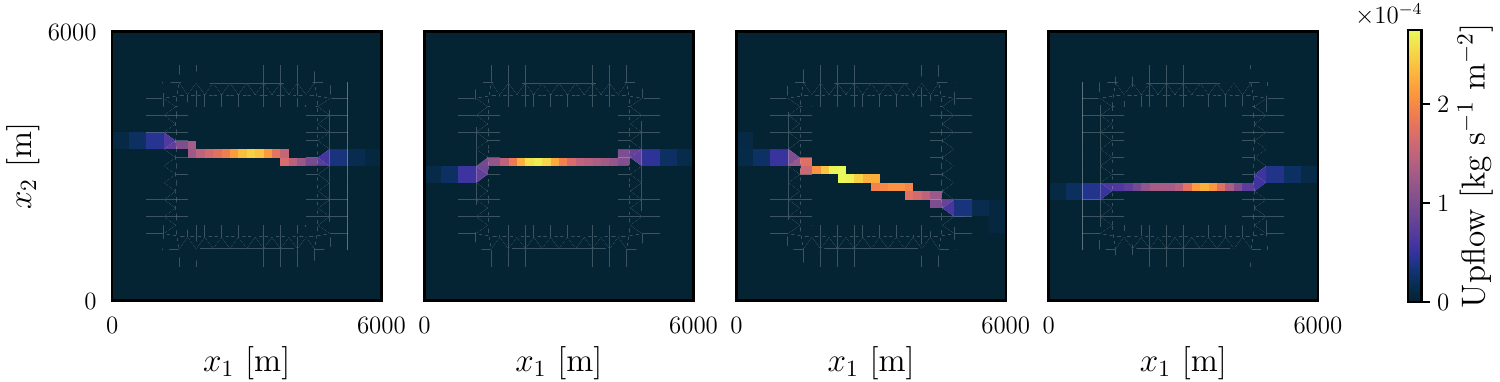}
    \caption{Mass upflows sampled from the prior.}
    \label{fig:geo_upflows}
\end{figure}

We assume that there is a single linear, vertical fault running through the reservoir from east to west. However, we treat the exact location of the fault, as well as the magnitude of the mass upflow contained within it, as unknown. We assume that the points at which the fault intersects the eastern and western boundaries of the model domain are independent and uniformly distributed between 1500\,m and 4500\,m, and that the upflow within each cell along the fault is modelled using a Gaussian random field with a squared-exponential covariance function (see Eq.~\ref{eq:sq_exp}) and a mean and standard deviation that reduce as the horizontal distance to the centre of the model domain increases, reflecting a prior belief that the upflow tends to be greatest near the centre of the model domain. The parameters of this GRF are chosen such that the total mass upflow entering the model domain tends to be between 80\,kg\,s$^{-1}$ and 120\,kg\,s$^{-1}$. Figure \ref{fig:geo_upflows} shows several samples of mass upflows drawn from the prior.

\begin{figure}
    \centering
    \includegraphics[width=0.95\textwidth]{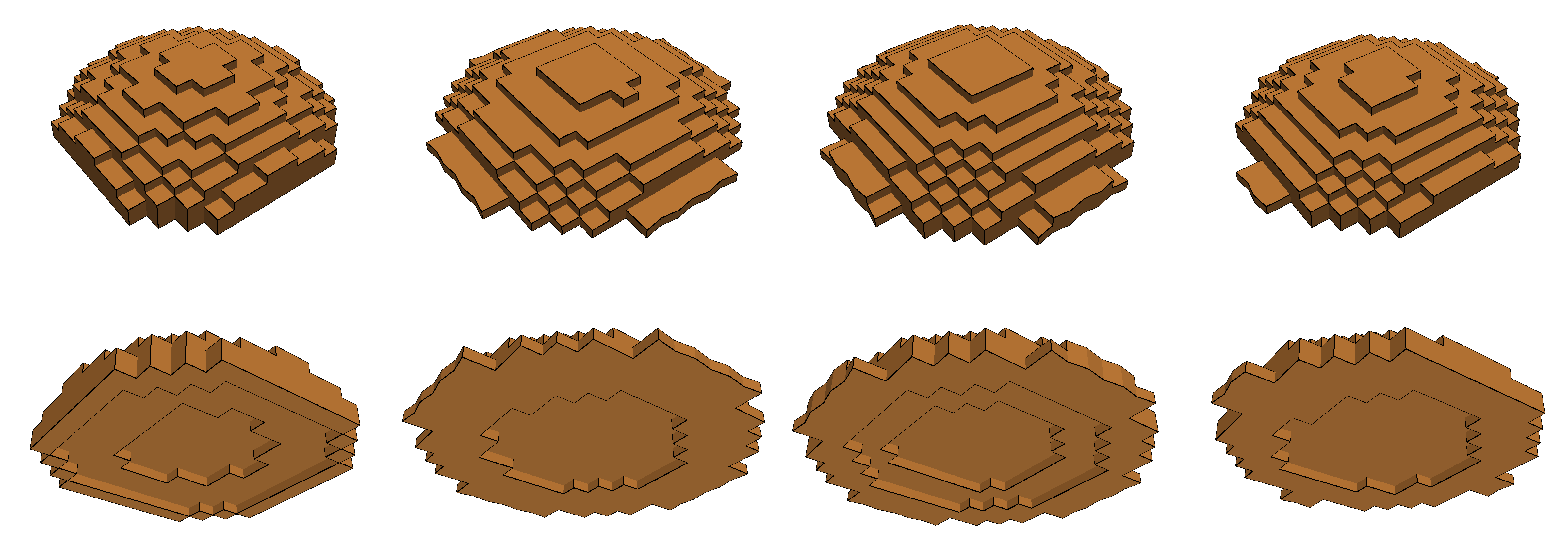}
    \caption{The top surfaces (top row) and bottom surfaces (bottom row) of clay cap geometries sampled from the prior.}
    \label{fig:geo_caps}
\end{figure}

To parametrise the permeability and porosity of the reservoir, we first partition the domain of the model into three regions with variable interfaces: the fault (which has a high permeability and porosity), a clay cap (which has a low permeability and porosity), and a background region (which has a moderate permeability and porosity). We model the clay cap as the deformation of a star-shaped set, with a boundary represented using a truncated Fourier series with uncertain coefficients. Figure \ref{fig:geo_caps} shows several clay cap geometries drawn from the prior; for a complete description of this parametrisation, the reader is referred to \citet{de2024ensemble}. Where the clay cap and fault intersect, the clay cap takes priority.

\begin{figure}
    \centering
    \includegraphics[width=0.95\textwidth]{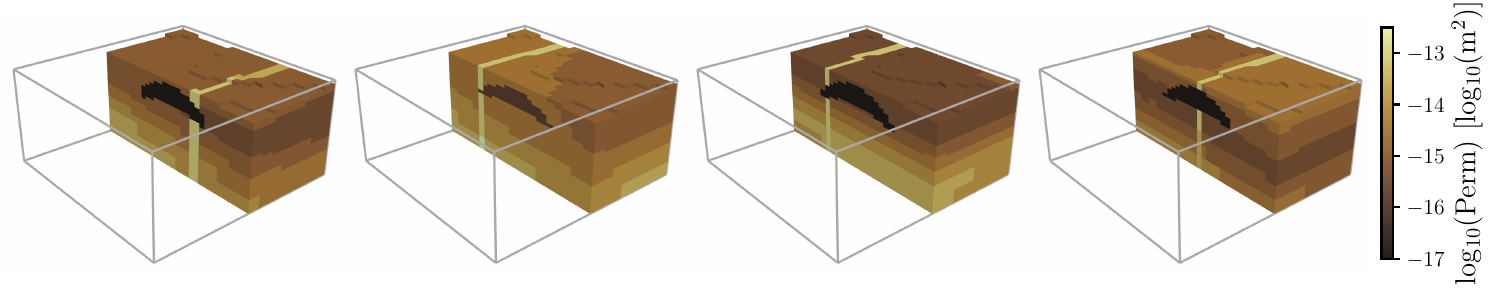}
    \caption{Permeability structures sampled from the prior.}
    \label{fig:geo_samples}
\end{figure}

To model the permeability and porosity within each region, we use the level set method \citep{iglesias2016bayesian}, which is often used in the modelling of subsurface systems to generate distinct regions with common geophysical characteristics \citep[see, e.g.,][]{muir2020geometric, tso2021efficient}. These regions are defined using the contours of a continuous function, referred to as the level set function, which we denote using $\varphi(\cdot)$. In each region, we first specify a set of rock types with varying permeabilities and porosities, then select a set of constants at which to threshold the level set function to produce each rock type. For instance, the permeability and porosity within the fault region are given by 
\begin{equation}
    (\Perm(\bs{x}), \phi(\bs{x})) = \begin{cases}
        (10^{-13.5}\,\mathrm{\preddim}^{2}, 0.20), & \phantom{+}\varphi(\bs{x}) < -0.5, \\
        (10^{-13.0}\,\mathrm{\preddim}^{2}, 0.25), & -0.5 \leq \varphi(\bs{x}) < 0.5, \\
        (10^{-12.5}\,\mathrm{\preddim}^{2}, 0.30), & \phantom{+}0.5 \leq \varphi(\bs{x}).
    \end{cases}
\end{equation}
The permeability and porosity of each of the other regions is defined similarly. We allow the permeability of the clay cap to vary between $10^{-17}\,\mathrm{\preddim}^{2}$ and $10^{-16}\,\mathrm{\preddim}^{2}$, and the porosity to vary between $0.05$ and $0.10$. We allow the permeability of the background region to vary between $10^{-15.5}\,\mathrm{\preddim}^{2}$ and $10^{-13.5}\,\mathrm{\preddim}^{2}$, and the porosity to vary between $0.10$ and $0.20$. In all regions, we choose the level set function to be a centred GRF with a squared-exponential covariance function, with a standard deviation chosen such that the prior probabilities of a given location within each region belonging to each rock type are approximately equal. The lengthscale of each level set function in the horizontal ($x_{1}$ and $x_{2}$ directions) is 8000\,m, while the lengthscale in the vertical ($x_{3}$) direction is 2000\,m. This tends to result in the generation of layered structures. Figure \ref{fig:geo_samples} shows several permeability structures sampled from the prior.

\subsubsection{Data}

\begin{figure}
    \centering
    \includegraphics[width=0.75\textwidth]{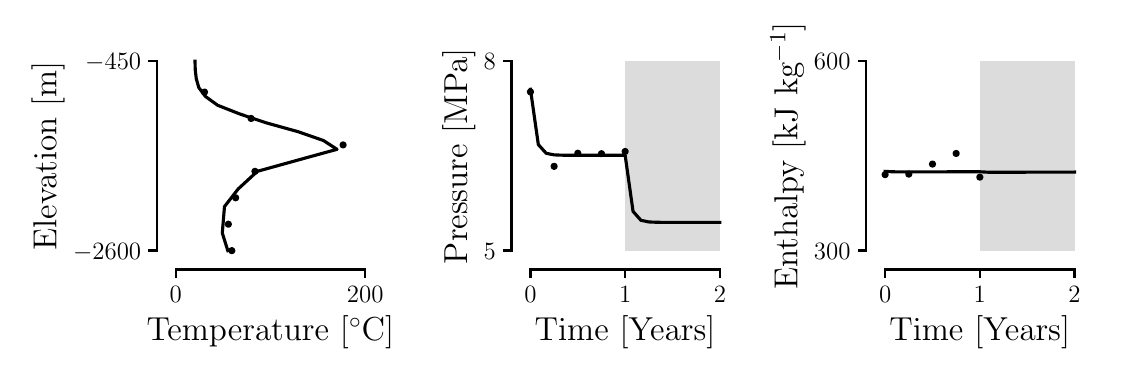}
    \caption{The natural state downhole temperature data (left), transient pressure data (centre) and transient enthalpy data (right) collected at well 2. In all plots, the solid line denotes the true reservoir state and the dots denote noisy observations. In the pressure and enthalpy plots, the grey region denotes the forecast period.}
    \label{fig:geo_data}
\end{figure}

We assume that we have access to measurements of the natural state temperature at seven equispaced points down each of wells 1--7, as well as measurements of the pressure and enthalpy of the fluid extracted from each well at three-month intervals over the first year of the production period. This gives a total of 119 measurements. We add independent Gaussian noise to each measurement, with a standard deviation equal to 2\% of the maximum value of the corresponding data type. Figure \ref{fig:geo_data} shows the data collected at well 1. No data is collected at well 8 or well 9.

\subsubsection{Simulation}

All simulations are carried out using the open-source simulator Waiwera \citep{croucher2020waiwera}, which uses a finite volume discretisation of the governing equations in Section \ref{sec:equations}, and are run in parallel on a high-performance computing cluster provided by New Zealand eScience Infrastructure. As in our previous model problem, to avoid the ``inverse crime'' of generating the synthetic data and solving the inverse problem using the same numerical discretisation \citep{kaipio2006statistical, kaipio2007statistical}, we use a mesh comprised of 13,383 cells when simulating the dynamics of the true system (generated using a draw from the prior), but a mesh comprised of 8788 cells (plotted in Figure \ref{fig:geo_mesh}) when carrying out the simulations required to approximate the posterior predictive distribution using DSI.

For the DSI algorithm, we run 2000 simulations using sets of parameters drawn from the prior. As is common in geothermal reservoir modelling, some of these converge to physically unreasonable values (for example, the reservoir pressure reduces below atmospheric pressure), while others do not converge entirely. After discarding these, we are left with 1413 simulations. We use 1300 of these as part of the DSI algorithm. Of the remaining samples, 100 are used to evaluate the quality of the DSI mapping; we elaborate on this further in the next section.

\subsubsection{Validation}

\begin{figure}
    \centering
    \includegraphics[width=0.95\linewidth]{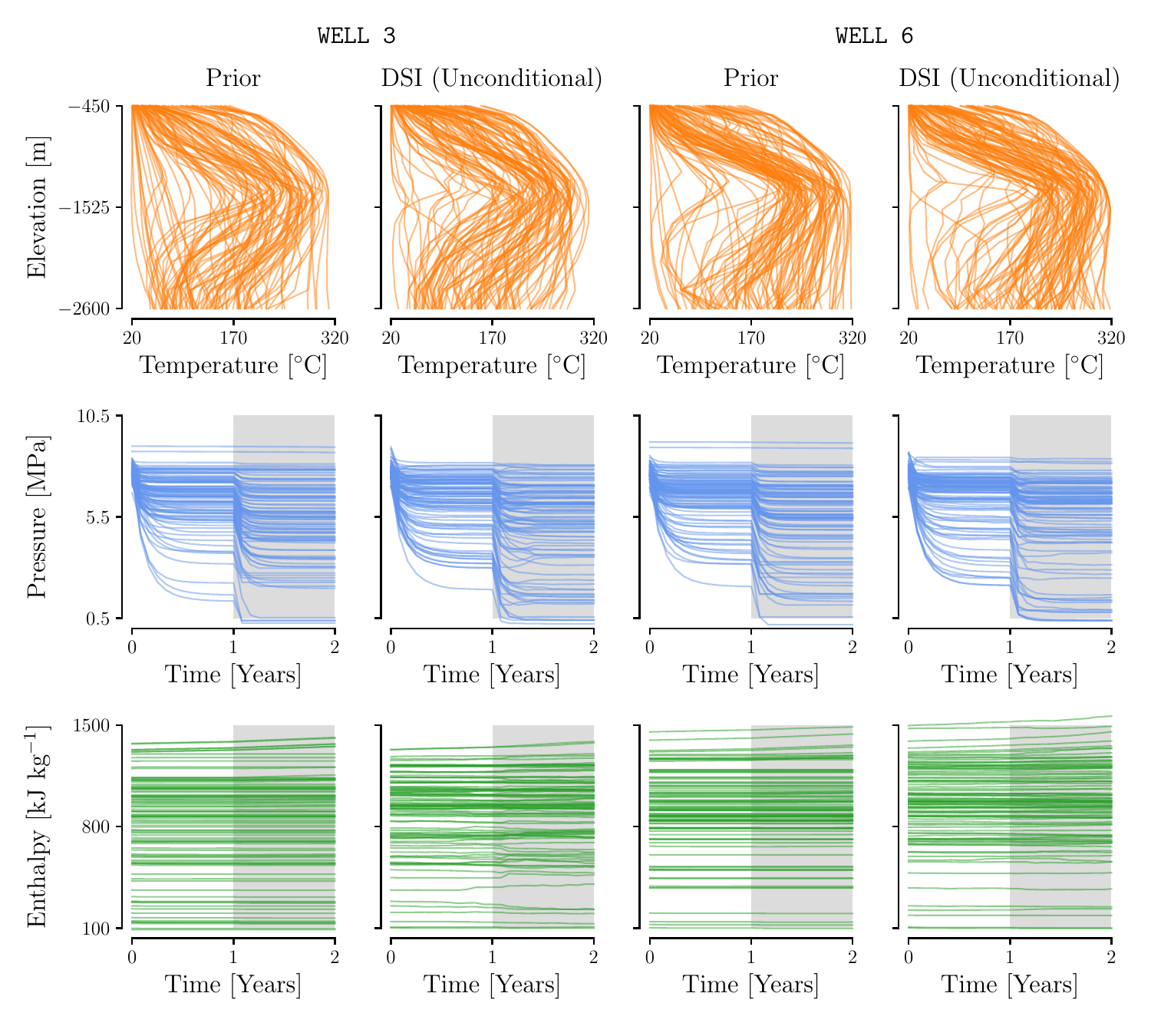}
    \caption{A set of 100 samples of the downhole temperatures at the end of the production history period, and the transient feedzone pressure and enthalpy, in well 6 (first and second columns) and well 8 (third and fourth columns). For both wells, samples from the prior predictive distribution and the DSI approximation to the prior predictive distribution are shown.}
    \label{fig:geo_prior_vs_dsi}
\end{figure}

As in the two-dimensional case study, we first evaluate the quality of the DSI mapping, $\mathcal{T}(\cdot)$, by plotting a set of (unconditional) realisations from the DSI approximation to the prior predictive distribution of the downhole temperature profiles at the end of the production history period, and the transient feedzone pressure and enthalpy of the system, at wells 6 and 8. These results are shown in Figure \ref{fig:geo_prior_vs_dsi}, which also functions as a prior predictive check. Again, we observe that the DSI predictions exhibit similar qualitative behaviour to the prior predictions.

As an additional form of validation, we evaluate the quality of the DSI surrogate using 100 of the samples from the prior predictive distribution of the data and predictive quantities of interest that were not used to construct it. For each of these ``validation'' datasets, we draw 1000 samples from the DSI approximation to the posterior predictive distribution. We then compute the proportion of each of the true values of the predictive quantities of interest (comprised of the downhole temperature profiles down each well at the end of the production history period, and the transient pressure and enthalpy at each feedzone over the forecast period) that are contained within the central 95 percent of the predictions. We find that, on average, 92.0 percent of each of the predicted temperature profiles, 93.2 percent of each of the predicted pressure profiles, and 92.3 percent of the predicted enthalpy profiles are contained within the central 95 percent of the predictions. This gives us confidence that the DSI algorithm is able to reduce our level of uncertainty in the values of these predictive quantities of interest without discounting the true values.

\subsubsection{Results}

\begin{figure}
    \centering
    \includegraphics[width=0.95\textwidth]{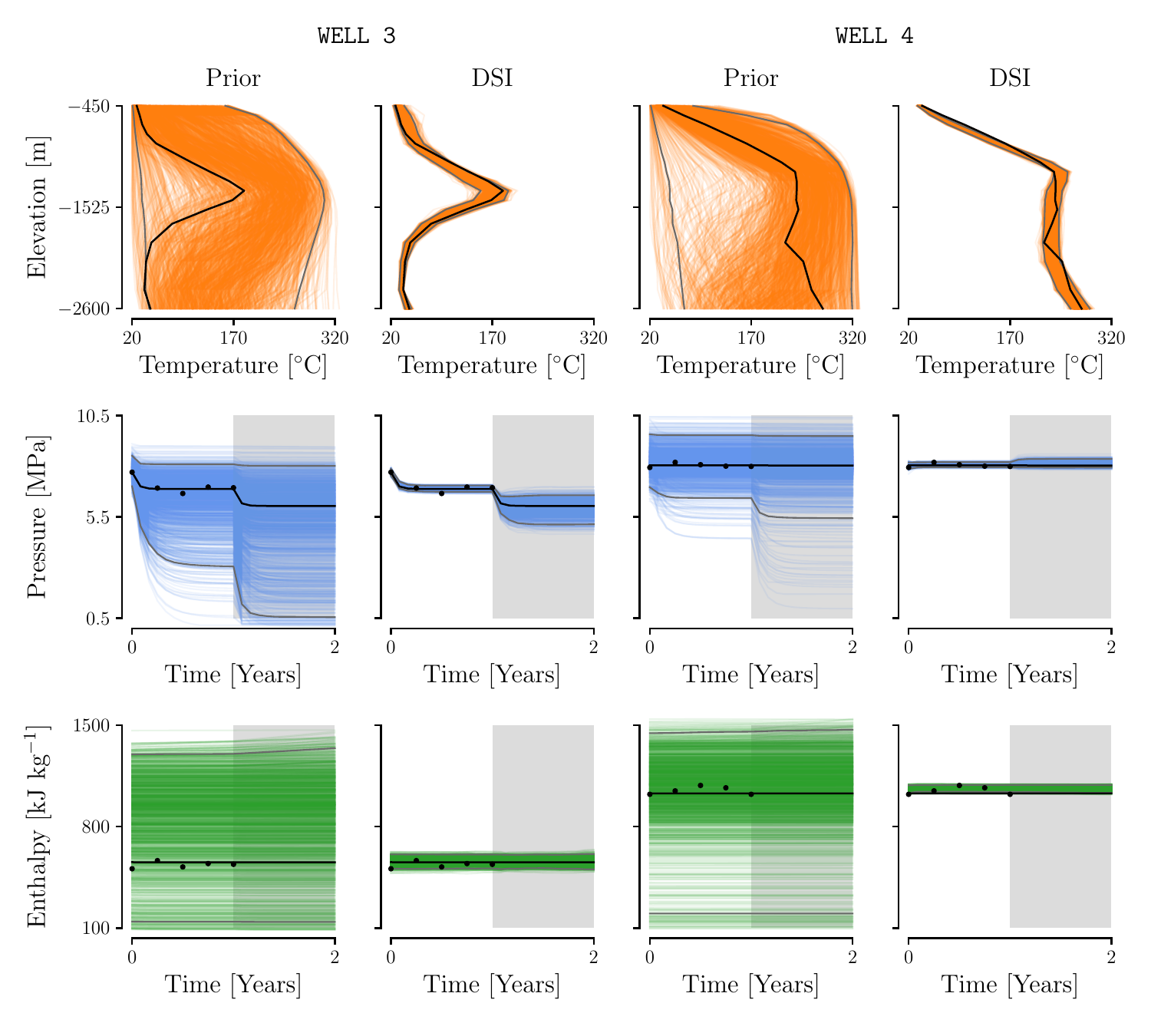}
    \caption{The downhole temperatures at the end of the production history period, and the transient feedzone pressure and enthalpy, in well 3 (first and second columns) and well 4 (third and fourth columns). For both wells, samples from the prior predictive distribution and the approximation to the posterior predictive distribution computed using DSI are shown. In all plots, the black line denotes the true reservoir state and the grey lines indicate the central 95\% of the samples. In the pressure and enthalpy plots, the black dots denote the noisy observations and the grey region denotes the forecast period.}
    \label{fig:geo_preds_34}
\end{figure}

We now discuss the results obtained when DSI is applied to the ``true'' system plotted in Figure \ref{fig:geo_truth}. Figure \ref{fig:geo_preds_34} shows samples from the prior predictive distribution, and the approximation to the posterior predictive distribution obtained using DSI, for the downhole temperature profiles at the end of the production history period, and the transient feedzone pressure and enthalpy over the production history and forecast periods, in well 3 and well 4. Results for other wells at which data is collected are similar. In all cases, applying the DSI algorithm gives a significant reduction in uncertainty, and the state of the true system generally has high posterior probability.

\begin{figure}
    \centering
    \includegraphics[width=0.95\textwidth]{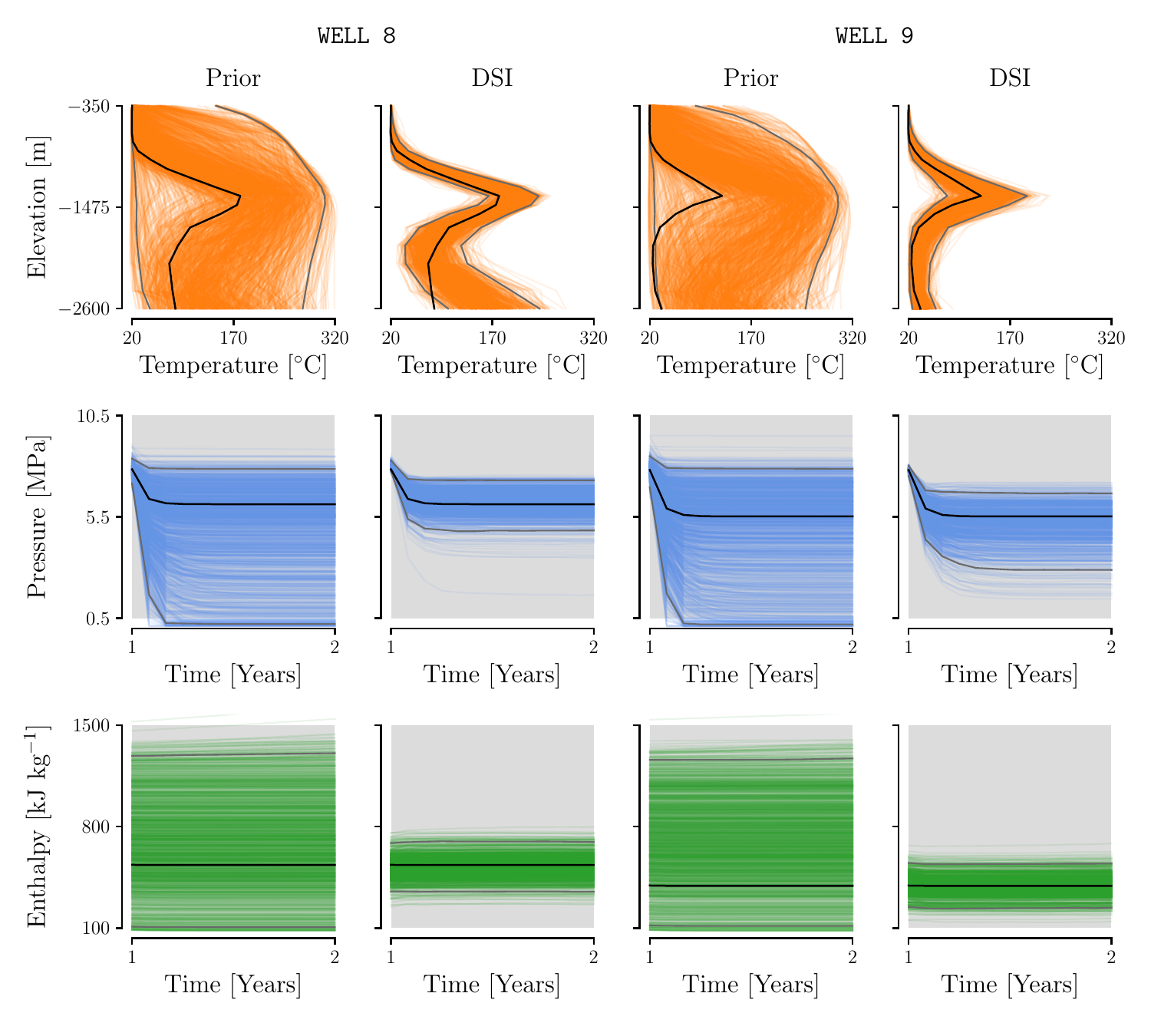}
    \caption{The downhole temperatures at the end of the production history period, and the transient feedzone pressure and enthalpy, in well 8 (first and second columns) and well 9 (third and fourth columns), over the forecast period. For both wells, samples from the prior predictive distribution and the approximation to the posterior predictive distribution computed using DSI are shown. In all plots, the black line denotes the true reservoir state and the grey lines indicate the central 95\% of the samples.}
    \label{fig:geo_preds_89}
\end{figure}

Figure \ref{fig:geo_preds_89} shows samples of the same quantities as Figure \ref{fig:geo_preds_34}, but for well 8 and well 9, at which no data is collected, and which only operate during the forecast period. In both cases, the uncertainty in each predictive quantity of interest is reduced after applying the DSI algorithm, and the state of the true system has high posterior probability. However, these reductions in uncertainty are significantly less than the corresponding reductions in uncertainty for well 3 and well 4; this is expected, given that we do not have access to any direct information on the state of the reservoir down these wells.

\begin{figure}
    \centering
    \includegraphics[width=0.95\textwidth]{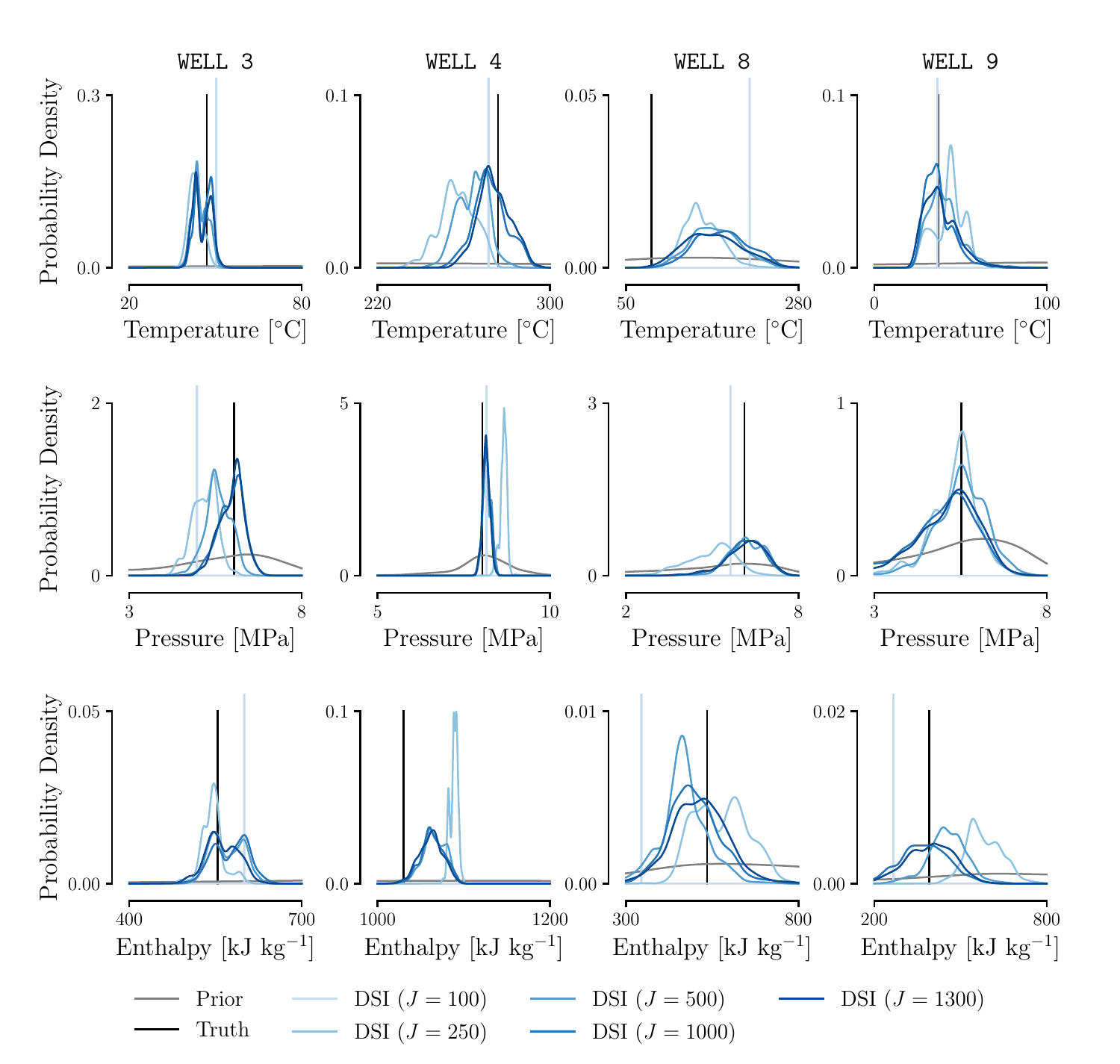}
    \caption{The prior predictive distribution and posterior predictive distributions of the temperature at the bottom of the well at the end of the production history period, and the feedzone pressure and enthalpy at the end of the forecast period, for wells 3, 4, 8 and 9. The black line in each plot denotes the true reservoir state.}
    \label{fig:geo_sample_comp}
\end{figure}

Finally, Figure \ref{fig:geo_sample_comp} shows how the posterior predictive distributions for the temperature at the bottom of each well at the end of the production history period, and the feedzone pressure and enthalpy at the end of the forecast period change as the number of samples used to estimate the mapping $\mathcal{T}(\cdot)$ varies, for wells 3, 4, 8 and 9. As in the two-dimensional setting, we observe that, when 100 samples are used, the resulting estimates appear significantly different to those obtained using a greater number of samples. Additionally, they often fail to capture the truth with non-negligible probability. When $500$ samples are used, the resulting estimates appear to be fairly consistent with those obtained using larger sample sizes, suggesting that $500$ or more samples is likely to be an appropriate number to use for this particular problem.

\section{Conclusions and Future Work}

In this work, we have introduced a simple variant of the data space inversion methodology that allows for efficient, approximate characterisation of the posterior predictive distribution. We have demonstrated that the resulting methodology can provide an efficient, derivative-free means of making geothermal reservoir model predictions, with quantified uncertainty, conditioned on observed data. We have also illustrated how the approximation to the posterior predictive distribution generated using DSI changes as the number of samples from the prior that are used as part of the algorithm is varied, and provided a systematic comparison of DSI and linearisation about the MAP estimate, another commonly-used technique for approximate Bayesian inference.

We emphasise that there are a number of limitations of the DSI framework. Like most surrogate modelling techniques, there exists little in the way of theoretical guarantees surrounding the quality of the DSI approximation to the posterior predictive distribution when applied to general inverse problems, and there is no guarantee that results obtained using DSI will respect the physics of the problem under consideration. There are also some elements of the DSI framework that warrant further study, including the sensitivity of the method to the choice of prior used, the dimension of the data, and the particular characteristics of the forward problem under consideration. Nonetheless, the empirical results we have obtained in this work appear promising.

An obvious next step will be to demonstrate the application of the DSI methodology to a real-world case study arising in geothermal reservoir modelling. A model of a real-world geothermal system is likely to require a more complex prior than we have used in our model problems; for instance, it is likely to need to account for the full, anisotropic permeability structure of the reservoir. There may also be additional data to consider, such as $\rm{CO}_{2}$ fractions or information on surface features. However, we anticipate that the application of the DSI framework will remain much the same.

Additionally, we have identified a variety of extensions of the DSI framework that would be valuable to explore in a geothermal setting. First, it is worth noting that while it is generally predictive quantities that are of most interest in a geothermal setting, it is also possible to use DSI to approximate the solution to the calibration problem (see Sec.~\ref{sec:calibration}); that is, the problem of estimating the model parameters. This simply amounts to using the samples of the parameters instead of the samples of the predictive quantities of interest when building the DSI surrogate; the process of drawing samples from the resulting approximation to the (parameter) posterior remain the same. We note that these ideas are used in the work of \citet{park2020direct}, which employs elements of the Bayesian evidential learning framework. It would be of interest to apply ideas from DSI to estimate the parameters of a geothermal reservoir model, and to compare the results obtained to those generated using other surrogate modelling approaches \citep[see, e.g.,][]{han2024surrogate, han2025accelerated}.

Additionally, the DSI framework is likely to be well-suited to solving optimal experimental design problems, in which one is interested in identifying a data collection plan that minimises a measure of the expected posterior uncertainty in the model parameters or predictive quantities of interest \citep{alexanderian2021optimal}. Solving an OED problem generally requires the repeated computation of the parameter posterior or posterior predictive distribution associated with sets of possible data one could expect to collect; because the DSI framework uses the same set of simulations from the prior when approximating these distributions, regardless of the data collected, solving an OED problem would require no more reservoir model simulations than solving a single inverse problem. For similar ideas within the context of the Bayesian evidential learning framework, see \citet{thibaut2021new}. 

A final area of interest is optimal control. The application of DSI to optimal control problems is studied in \citet{jiang2020data}, which uses an extension of the DSI framework to optimise the management of oil reservoirs, by treating user-specified well controls as ``data'' to be conditioned. This allows for the efficient approximation of the posterior predictive distribution under various management scenarios.

Overall, we believe that our results demonstrate that data space inversion has the potential to be a useful tool in geothermal reservoir modelling, and should be investigated further.

\section*{Acknowledgments}

The authors wish to acknowledge the use of New Zealand eScience Infrastructure (NeSI; \url{www.nesi.org.nz}) high performance computing facilities as part of this research. These facilities are funded jointly by NeSI's collaborator institutions and through the Ministry of Business, Innovation \& Employment's Research Infrastructure programme. The authors would like to thank the MBIE research programme ``Empowering Geothermal'' which has in part funded this research. Finally, the authors wish to thank John Doherty for several fruitful discussions on the data space inversion methodology, and the two anonymous reviewers, whose comments have improved this paper significantly.

\section*{Open Research Section}

The code and models used to carry out the experiments and generate the figures in this paper are archived on Zenodo \citep{de2024}, and are also available on GitHub (\url{https://github.com/alexgdebeer/GeothermalDSI}) under the MIT license.

\renewcommand*{\mkbibnamefamily}{\textsc}
\printbibliography

\end{document}